\documentclass[aps,prc,twocolumn,superscriptaddress,showpacs]{revtex4}
\usepackage{graphicx}
\usepackage{dcolumn}
\usepackage{bm}
\usepackage{subfigure}
\usepackage{amssymb}
\usepackage{amsmath}
\newcommand{\beq}{\begin{equation}}
\newcommand{\eeq}{\end{equation}}
\newcommand{\beqar}{\begin{eqnarray}}
\newcommand{\eeqar}{\end{eqnarray}}
\newcommand{\ds}{\displaystyle}
\usepackage{color}

\begin{document}

\title{ Shear viscosity in microscopic calculations of {\it A+A} collisions 
at energies available at the Nuclotron-based Ion Collider fAcility (NICA) }

\author{M.~Teslyk}
\altaffiliation[Also at ]{
Taras Shevchenko National University of Kyiv, UA-01033 Kyiv, Ukraine
\vspace*{1ex}}
\affiliation{
Department of Physics, University of Oslo, PB 1048 Blindern,
N-0316 Oslo, Norway
\vspace*{1ex}}
\author{L.~Bravina}
\affiliation{
Department of Physics, University of Oslo, PB 1048 Blindern,
N-0316 Oslo, Norway
\vspace*{1ex}}
\author{O.~Panova}
\altaffiliation[Also at ]{
Taras Shevchenko National University of Kyiv, UA-01033 Kyiv, Ukraine
\vspace*{1ex}}
\affiliation{
Department of Physics, University of Oslo, PB 1048 Blindern,
N-0316 Oslo, Norway
\vspace*{1ex}}
\author{O.~Vitiuk}
\altaffiliation[Also at ]{
Taras Shevchenko National University of Kyiv, UA-01033 Kyiv, Ukraine
\vspace*{1ex}}
\affiliation{
Department of Physics, University of Oslo, PB 1048 Blindern,
N-0316 Oslo, Norway
\vspace*{1ex}}
\author{E.~Zabrodin}
\altaffiliation[Also at ]{
Skobeltsyn Institute of Nuclear Physics,
Moscow State University, RU-119991 Moscow, Russia
\vspace*{1ex}}
\affiliation{
Department of Physics, University of Oslo, PB 1048 Blindern,
N-0316 Oslo, Norway
\vspace*{1ex}}

\date{\today}

\begin{abstract}
Time evolution of shear viscosity $\eta$, entropy density $s$, and their 
ratio $\eta / s$ in the central area of central gold-gold collisions at
energies available at the Nuclotron-based Ion Collider fAcility (NICA) 
are studied within the ultrarelativistic quantum molecular dynamics
(UrQMD) transport model. The extracted values of energy density, net 
baryon density, and net strangeness density are used as input to (i) the
statistical model of an ideal hadron gas to define temperature, 
baryochemical potential, and strangeness chemical potential, and to 
(ii) a UrQMD box with periodic boundary conditions to study the 
relaxation process of highly excited matter. During the relaxation stage, 
the shear viscosity is determined in the framework of the Green-Kubo 
approach. The procedure is performed for each of 20 time slices, 
corresponding to conditions in the central area of the fireball at times 
from 1 to 20~fm/$c$. For all tested energies the ratio $\eta / s$ 
reaches its minimum, $\ds \left( \eta/s \right)_{min} \approx 0.3$ at 
$t \approx 5$~fm/$c$. Then it increases up to the late stages of the 
system evolution. This rise is accompanied by the drop of both 
temperature and strangeness chemical potential and by the increase of 
baryochemical potential.
\end{abstract}

%
%
\maketitle

\section{Introduction}
\label{sec:intro}

Relativistic heavy-ion collisions have been intensively studied both 
theoretically and experimentally to obtain information about the 
properties of highly excited nuclear matter. To date, these collisions
are the only means to study the conditions of early Universe in the
laboratory, thus leading to the term ``little big bang" \cite{Shur_04}.
According to the theoretical estimates and lattice quantum 
chromodynamics (lQCD) calculations, nuclear matter under certain extreme 
conditions should experience a deconfinement phase transition into the a
phase of matter, a quark-gluon plasma (QGP). Expanding hot fireball 
should, however, rapidly cool off, and the plasma will undergo 
hadronization. Experiments show that in heavy-ion collisions at the 
ultrarelativistic energies of the Relativistic Heavy Ion Collider (RHIC), 
$\sqrt{s} = 200$~GeV, and of the Large Hadron Collider (LHC), $\sqrt{s}= 
2.76$ and 5.02~TeV, there is a crossover type of the phase transition. 
In contrast, at much lower energies the transition might be of the first 
order. In this case the line of the first-order phase transition in the
nuclear phase diagram ends up in the tricritical point, where the 
transition becomes of second order. The search for the tricritical point 
is in the agenda of experiments with heavy-ion beams at the forthcoming
Nuclotron-based Ion Collider fAcility (NICA) and the Facility for 
Antiproton and Ion Research (FAIR), and within the beam energy scan 
(BES) program at RHIC. Therefore, one has to look for the observable 
most sensitive to the QGP--hadrons transition. One such observable is 
the ratio of shear viscosity $\eta$ to entropy density $s$, $\eta/s$. 
This ratio drops to a minimum at critical temperatures for all known 
substances \cite{CKML_06}, and in relativistic heavy-ion collisions it 
is expected to be of order of its theoretical lower bound, $1/4\pi$ 
\cite{KSS_05}; for details see, e.g., \cite{PRL.99.172301} and 
references therein.

Despite the interest in this topic, it is still difficult to estimate 
the value of the ratio exactly due to the high calculation complexity 
required by QCD simulations. Therefore, various works in the field have 
explored different approaches and approximations for the conditions 
expected to prevail near the phase transition; see, e.g., 
\cite{PRC.69.044901,DB_09,CK_11,PRC.84,MPLB.25,PRC.86,PRC.86.054902,PRC.87,PRC.91,IS_16}. 
For example, in \cite{PRC.69.044901} thermodynamic quantities of 
hadronic matter are studied for a system of light mesons embedded in a 
box with periodic boundary conditions generated by the ultrarelativistic 
quantum molecular dynamics (UrQMD) model. A relativistic hadron gas in 
thermal and chemical equilibrium and with zero baryon and strangeness 
chemical potentials was considered in \cite{DB_09}.
In \cite{PRC.86.054902} the authors obtain viscosity $\eta$ by solving 
the ultrarelativistic Boltzmann transport equation and compare it to the 
one obtained via the Chapman-Enskog approximation. Recently, the shear 
viscosity and its ratio to entropy density were calculated for a gas of 
Hagedorn states \cite{RGC_19} with masses up to 10~GeV/$c^2$. It was 
found that, because of the rapid growth of $s$ in the vicinity of 
Hagedorn limiting temperature $T_{\rm H} = 165$~MeV, the ratio $\eta /s$ 
came close to and was even below the bounding $1/4 \pi$ given by the
anti-de Sitter and conformal field theory (AdS-CFT) \cite{KSS_05}. Among 
the other papers on the topic are \cite{PRC.97.055204}, where viscosity 
is extracted within the SMASH transport model, and \cite{JPG_18}, where 
the UrQMD model was employed for system of nucleons at intermediate
temperatures between 10 and 50~MeV. In the latter case the nucleons 
were allowed to experience only elastic collisions.

Definitely, heavy-ion collisions at energies of NICA and higher are more
complex. As mentioned in \cite{PRC.78.024902}, the ratio $\eta/s$ 
cannot be constant during the evolution of the fireball. To provide 
better fits to the experimental data, this ratio should depend on both
temperature and chemical potentials. Consequently, it is essential to 
explore the time dynamics of the ratio $\eta/s$ from the very beginning 
of a relativistic heavy-ion collision.

In the present paper we investigate fluctuation relaxation time $\tau$ 
and shear viscosity $\eta$, as well as its ratio to entropy density 
$\eta/s$, for central Au+Au collisions calculated in the UrQMD model 
\cite{urqmd_1,urqmd_2} within the NICA energy range. Compared to the 
previous researches, we study the evolution of $\eta, s$, and $\eta/s$
in heavy-ion collisions, where all characteristics are quickly changing,
and not, e.g., the temperature dependence of the $\eta/s$ ratio at 
constant chemical potentials. Investigation of dynamics of the 
relaxation process in a box with periodic boundary conditions allows us 
to estimate both the lower and upper bounds of the time interval at 
different energies, where it is possible to extract $\tau$.

The paper is organized as follows. Section \ref{sec:model} describes
briefly the features of the UrQMD model and the UrQMD box calculations.
To extract the thermodynamic quantities, such as temperature $T$,
baryochemical potential $\mu_{\rm B}$, and strangeness chemical 
potential $\mu_{\rm S}$, one has to compare microscopic model 
calculations with the results provided by the statistical model (SM) of 
an ideal hadron gas with essentially the same degrees of freedom. This 
model is also explained in Sec.~\ref{sec:model}. The formalism employed 
to determine the shear viscosity of hot and dense nuclear matter is 
presented in Sec.~\ref{sec:method}. Section~\ref{sec:results} contains 
results of our study, including the time evolution of $\eta$ and 
$\eta/s$ in the central area of heavy-ion collisions, and dependencies 
of $\eta/s$ on $T$, $\mu_{\rm B}$, and $\mu_{\rm S}$. Finally, 
conclusions are drawn in Sec.~\ref{sec:conclusions}.

\section{Models employed for the analysis}
\label{sec:model}

In our study of shear viscosity we employ three computational models.
The first one is the microscopic transport model UrQMD to calculate 
{\it A+A} collisions at a given energy and get the bulk characteristics 
of hot and dense nuclear matter, namely, energy density $\varepsilon$, 
net baryon density $\rho_{\rm B}$, and net strangeness density 
$\rho_{\rm S}$. The second model is the UrQMD box with periodic boundary 
conditions to study the relaxation process and find the relaxation time 
$\tau$. Finally, to determine thermodynamic parameters of the 
equilibrated system, i.e., temperature $T$, baryon chemical potential 
$\mu_{\rm B}$, and strangeness chemical potential $\mu_{\rm S}$, we 
apply the statistical model of an ideal hadron gas. The main features of 
all three models are as follows.

\subsection{UrQMD model} 
\label{ssec:model_1}

This is a well-known model \cite{urqmd_1,urqmd_2}
widely used for the analysis of heavy-ion collisions in a broad energy 
range. UrQMD is based on covariant propagation of hadrons on classical
trajectories, stochastic binary interactions of these hadrons if the 
distance between them is less than $ d \leq d_0 = \sqrt{\sigma^{tot} / 
\pi}$, where $\sigma^{tot}$ is the total cross section, formation and 
decay of resonances, and, when a certain collision energy limit is 
exceeded, formation and subsequent fragmentation of specific colored
objects, strings. For the treatment of strings UrQMD employs classical
Lund model \cite{lund}. As independent degrees of freedom the model 
considers 55 different baryon states with masses up to $m_{\rm B}^{max}
\leq 2.25$~GeV/$c^2$ and 39 different meson states, including the 
charmed ones. The list of particles is supplemented by corresponding 
antiparticles and isospin-projected states. Cross sections of 
hadron-hadron ($hh$) interactions are taken from the available 
experimental data \cite{pdg}. If this information is missing, the model 
relies on the unitarity, the additive quark model, and detailed balance 
considerations.

\subsection{Calculation of nuclear infinite matter: UrQMD box} 
\label{ssec:model_r21}

The box with finite volume and periodic boundary 
conditions serves to simulate the properties of infinite nuclear matter
\cite{prc_98,prc_00}. All particle interactions assumed in UrQMD
are allowed in the box as well. However, if any particle leaves the box,
another particle with absolutely identical parameters enters it, thus 
ensuring the preservation of initial energy density, net baryon density,
and net strangeness density in the box. The initial state in the box can 
be generated as mixture of baryons and antibaryons, or a baryon-free gas
of mesons, or even a system of strings and resonances. In the case of 
nonzero net baryon charge and zero net strangeness it is convenient to 
initialize the box containing neutrons and protons only. All nucleons 
can be uniformly distributed in the space, whereas their momenta are 
randomly distributed in a Fermi sphere and then rescaled to ensure the 
required energy density. Note also that relaxation to equilibrium in 
the box proceeds much longer compared to, e.g., that in the central cell 
in heavy-ion collisions \cite{prc_00}. In an open-system-like cell, 
the most energetic particles leave it earlier, and the whole system is 
cooling down. In a closed-system-like box, one has to wait until the 
kinetic energy of the most ``hot" particles will be redistributed among 
other particles and also converted to the mass of newly produced hadrons.

Finally, we have to determine temperature and chemical potentials in the
system. This is done by multiple fits of hadron abundances and energy 
spectra in UrQMD to those calculated within the statistical model.

\subsection{Statistical model of ideal hadron gas.}    
\label{ssec:model_3}

If the system of hadrons containing $1 \leq i \leq n$ different species
is in equilibrium at temperature $T$, all many-particle correlations in
it are reduced to a set of distribution functions (in system of natural
units $c=\hbar = k_B = 1$)
\beq \ds
\displaystyle
f(p,m_i) = \left[ \exp{ \left( \frac{\epsilon_i - \mu_i}{T} \right) }
+ C \right] ^{-1}
\label{eq1}
\eeq
Here $C = +1$ for fermions and $C = -1$ for baryons, and $p, m_i, 
\epsilon_i,$ and $\mu_i$ are hadron momentum, mass, 
energy, and chemical potential, respectively. The last depends on 
chemical potentials assigned to baryon charge $B_i$, strangeness content
$S_i$, and electric charge $Q_i$ of $i$-th hadron. However, the chemical
potential $\mu_{\rm Q}$ of electric charge is usually much smaller 
compared to baryochemical potential $\mu_{\rm B}$ and strangeness
chemical potential $\mu_{\rm S}$. Therefore, we will consider the linear
combination of two terms for the full chemical potential of a hadron:
\beq \ds
\mu_i = B_i \mu_{\rm B} + S_i \mu_{\rm S}
\label{eq2}
\eeq
The partial number density $n_i$ , the energy density $\varepsilon_i$, 
and the entropy density $s_i$ read
\beqar \ds
\label{eq3}
n_i &=&\frac{g_i}{2\pi^2}\int \limits_0^\infty f(p,m_i) p^2 dp \\
\label{eq4}
\varepsilon_i &=& \frac{g_i}{2\pi^2} \int \limits_0^\infty 
\sqrt{p^2+m_i^2} f(p,m_i) p^2 dp \\
\label{eq5}
s_i &=& - \frac{g_i}{2\pi^2} \int \limits_0^\infty
f(p,m_i) \left[ \ln{f(p,m_i)}-1 \right] p^2 d p \ ,
\eeqar
where $g_i$ is the spin-isospin degeneracy factor. The values of $T, 
\mu_{\rm B}$ and $\mu_{\rm S}$ should satisfy the set of nonlinear 
equations
\beqar \ds
\label{eq6}
\varepsilon &=& \sum_i \varepsilon_i (T, \mu_{\rm B}, \mu_{\rm S}) \\
\label{eq7}
\rho_B &=& \sum_i B_i\, n_i (T, \mu_{\rm B}, \mu_{\rm S}) \\
\label{eq8}
\rho_S &=& \sum_i S_i\, n_i (T, \mu_{\rm B}, \mu_{\rm S}) \ ,
\eeqar 
where $\varepsilon, \rho_{\rm B}$, and $ \rho_{\rm S}$ are taken as input
from microscopic model calculations.
 
\section{Shear viscosity determination procedure}
\label{sec:method}

We calculate central Au+Au collisions in the laboratory frame at energies 
$E_{lab} = 10A, 20A, 30A,\ {\rm and}\ 40${\it A}GeV, corresponding to 
$\sqrt{s}$ from 4.5 to 8.8~GeV in the center-of-mass frame. From the 
whole system the central cell with volume $ 5\times 5\times 5 =
125$~fm$^3$ is selected. Then, the energy density $\varepsilon$, the 
net baryon density $\rho_\mathrm{B}$, and the net strangeness density 
$\rho_\mathrm{S}$ in the cell are extracted at times $t_\mathrm{cell} = 
1 - 20$~fm/$c$ with the time step of 1~fm/$c$. In order to minimize 
statistical errors an ensemble of 51200 Au+Au central collisions at each 
energy has been generated.

The extracted data are inserted in the statistical model of the ideal 
hadron gas to obtain temperature $T$, entropy density $s_\mathrm{sm}$, 
baryon chemical potential $\mu_\mathrm{B}$, and strangeness chemical 
potential $\mu_\mathrm{s}$. After that we start UrQMD box calculations. 
The box with volume $V = 10 \times 10 \times 10 = 1000$~fm$^3 $ is 
initialized with the same values of $\varepsilon$, $\rho_\mathrm{B}$, 
and $\rho_\mathrm{S}$ as extracted from the cell analysis. Baryon 
density is provided by protons and neutrons taken in equal proportion, 
$N_p : N_n = 1 : 1$. Non-zero strangeness density is generated by the 
admixture of kaons. The box data are analyzed for times 
$t_\mathrm{box} = 1 - 1000$~fm/$c$ with the time step 1~fm/$c$. The box
ensemble consists of 12800 box simulations for each of 80 points.

To extract $\eta$ the Green-Kubo \cite{1.1740082,jpsj.12.570} formalism
was used. The formalism requires the existence of an equilibrated state 
in the medium in order to provide exponential damping of deviations from 
the equilibrium with time. Thus, the verification of equilibrium or of 
exponential damping of fluctuations is the necessary condition to be 
checked.

From the Green-Kubo formalism it follows that shear viscosity $\eta$ may 
be defined as
\beq \ds
\label{eta}
  \eta\left(t_0\right) = \frac{V}{T}\int_{t_0}^{\infty}
  \mathrm{d}t \langle \pi\left(t\right)\pi\left(t_0\right) \rangle_t \ ,
\eeq
where $t_0$ and $t$ denote moments of time in the box, and correlator 
$ \langle \pi\left(t\right)\pi\left(t_0\right) \rangle_t $ can be cast
in the form
\beqar \ds
\label{correlator}
\nonumber
  \langle \pi\left(t\right)\pi\left(t_0\right) \rangle_t &=& 
  \sum_{\substack{i,j=1 \\ i\neq j}}^{3}\frac{1}{3}\left[ \lim
  \limits_{t_\mathrm{max}\to\infty}\frac{1}{t_\mathrm{max}} 
\right. \\
& & \left. \times \int_{t_0}^{t_\mathrm{max}}\mathrm{d}t'\pi^{ij}
\left(t+t'\right) \pi^{ij}\left(t' \right) \right] 
\eeqar
with $\pi^{ij}$ being nondiagonal part of the stress-energy tensor
$T^{ij}$ 
\beq \ds
\label{pi^ij}
  \pi^{ij}\left(t\right) = \frac{1}{V}\sum_{i \neq j}
  \frac{p^i\left(t\right) p^j\left(t\right)}{E\left(t\right)} \ .
\eeq
Here $p^{i(j)}$ and $E$ are the $ i(j) $th components of momentum and 
energy of the particle, respectively. $ t_0 $ is the initial cut-off 
time indicating the beginning of the extraction of quantities from the 
box. The coefficient $1/3$ in the sum $\sum_{i,j} $ means averaging 
over the directions which allows one to reduce the statistical errors.
Usually the cutoff time $t_0$ is set to zero. We have left it here on 
purpose to explore the influence of the onset of data extraction from 
box calculations on the extracted value of shear viscosity.

If the system is in equilibrium, the correlator (\ref{correlator}) is 
expected to experience an exponential drop with time, i.e.,
\beq \ds
\label{correlator_exp}
  \langle \pi\left(t\right)\pi\left(t_0\right) \rangle_t = \langle 
  \pi\left(t_0\right)\pi\left(t_0\right) \rangle \exp\left(-\frac{t-t_0}
  {\tau}\right) \ ,
\eeq
with $\tau$ being an effective relaxation time of the system.

Inserting Eq.~(\ref{correlator_exp}) in Eq.~(\ref{eta}) one gets
\beq \ds
\label{eta tau}
  \eta\left(t_0\right) = \frac{\tau V}{T}\langle \pi
  \left(t_0\right)\pi\left(t_0\right) \rangle \ .
\eeq

As follows from Eq.~(\ref{eta tau}), the problem of evaluation of $\eta$ 
is reduced to estimation of $\tau$. Shear viscosity may be obtained then 
in two different ways: (i) by direct calculation of integral from 
Eq.~(\ref{eta}), which is equivalent to taking into account all time 
contributions to the correlator, or (ii) by fitting the correlator to 
Eq.~(\ref{correlator_exp}) in some selected time interval and applying 
Eq.~(\ref{eta tau}). The key difference here is the influence of 
fluctuations. The first case takes them into account and assumes that 
they are mostly mutually extinguished, whereas the second one cuts off 
fluctuations at times $t \gg \tau$ when the correlator is too small 
compared to the fluctuations (white noise) \cite{PRC.86.054902}. In 
what follows we compare the relaxation times $\tau$ for both cases.

\section{Results}
\label{sec:results}

\begin{figure}
\resizebox{\linewidth}{!}{
	\includegraphics[scale=0.60]{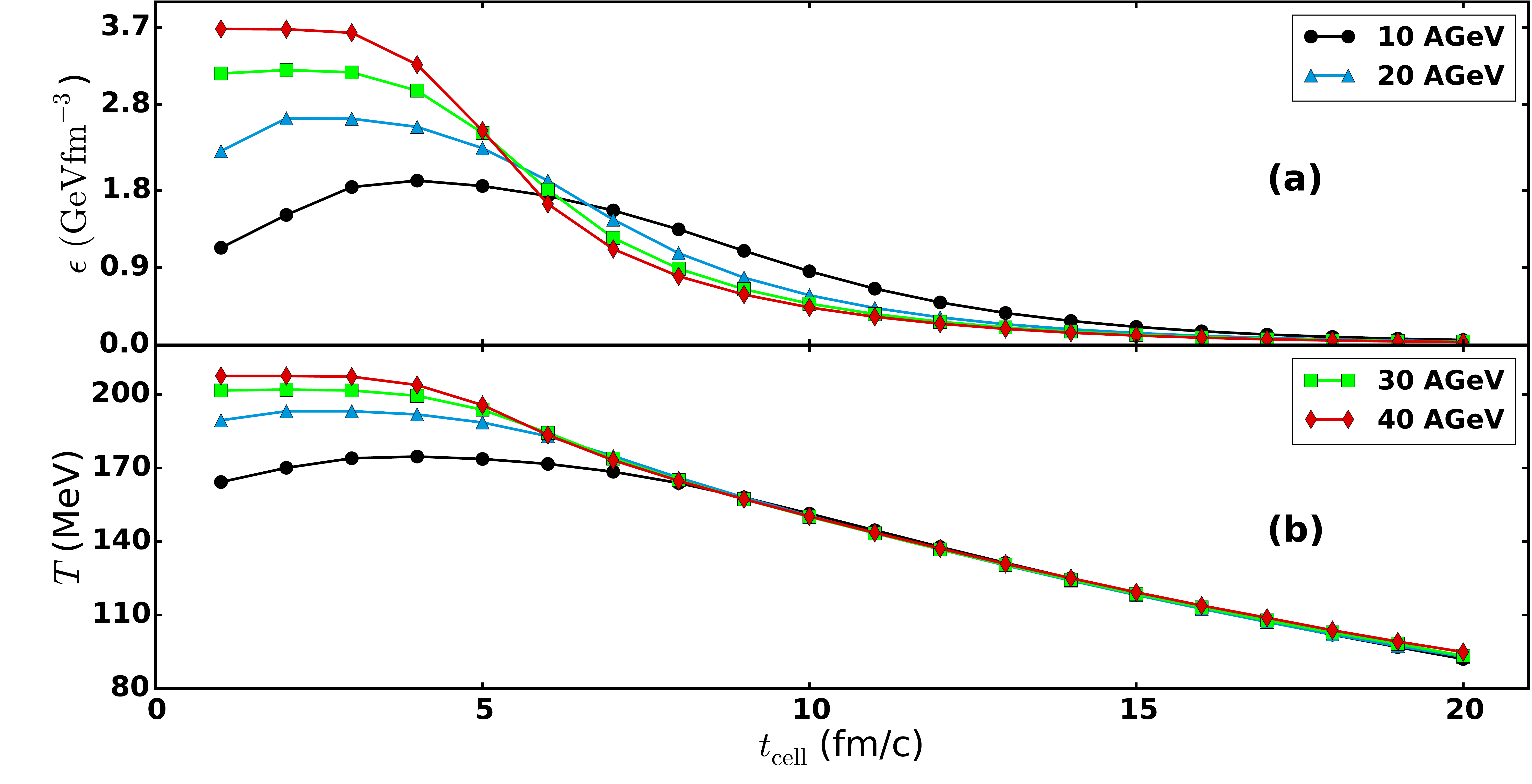}}
\resizebox{\linewidth}{!}{
	\includegraphics[scale=0.60]{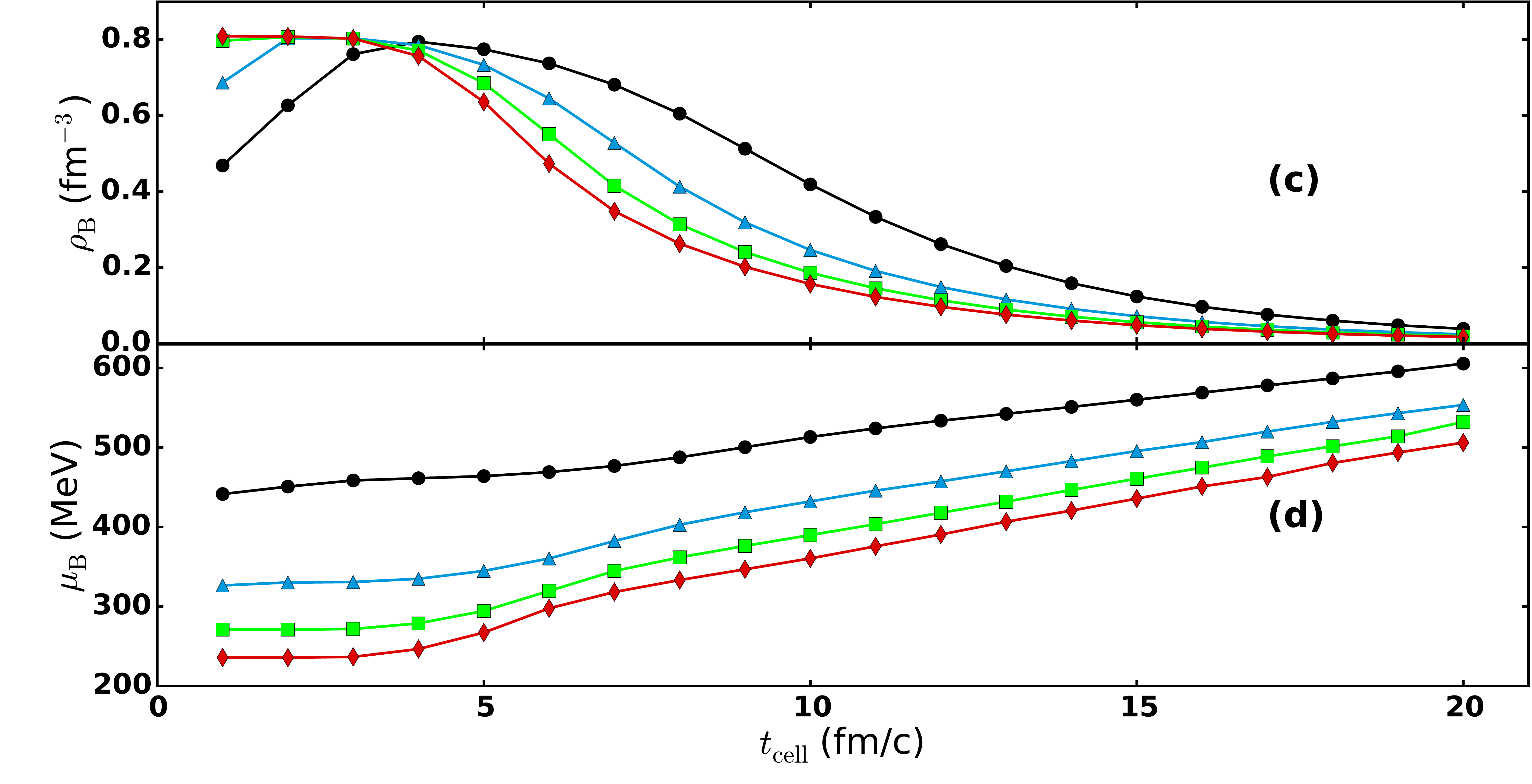}}
\resizebox{\linewidth}{!}{
	\includegraphics[scale=0.60]{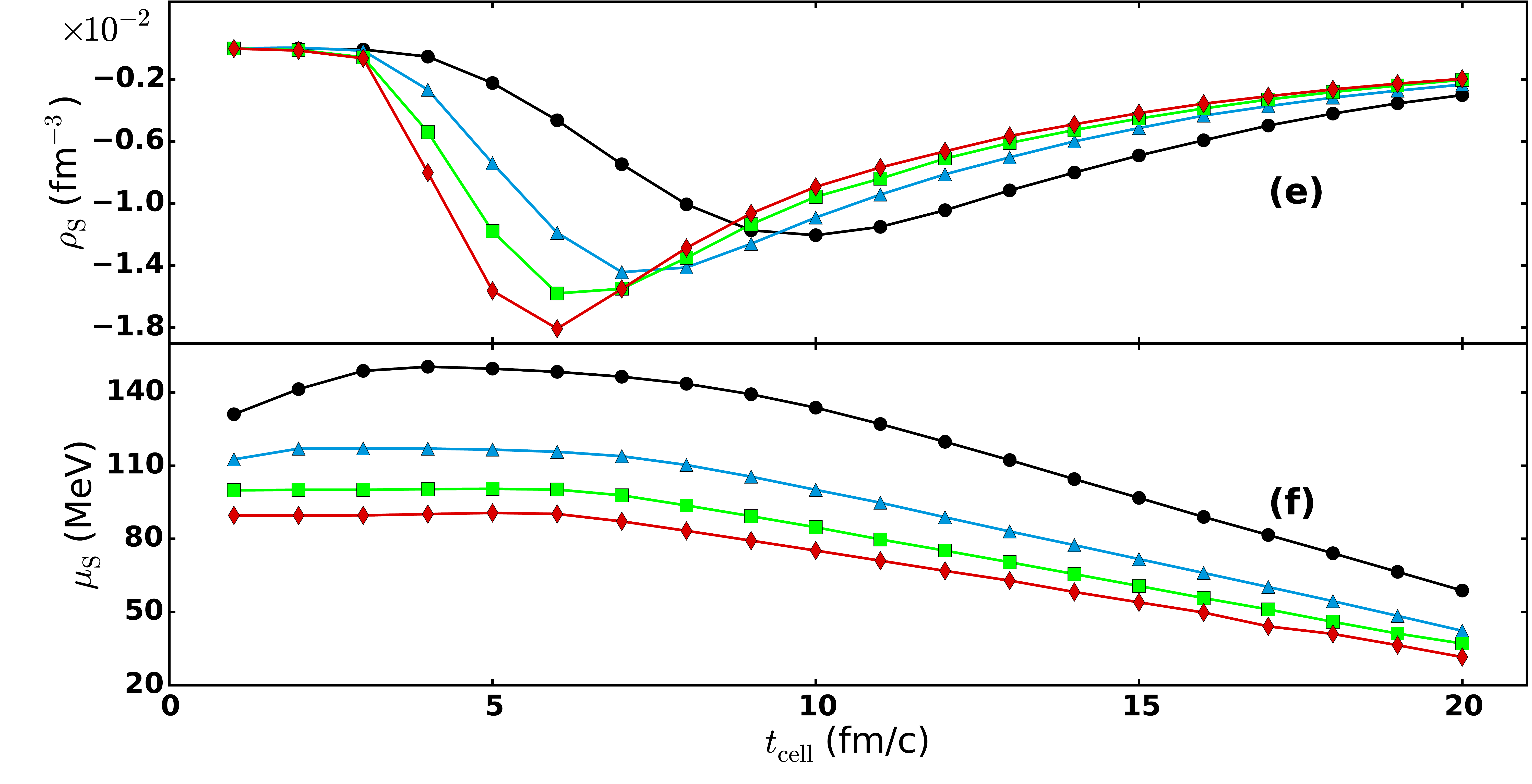}
}
\caption{(Color online)
Time evolution of (a) energy density $\varepsilon$, (c) net 
baryon density $\rho_\mathrm{B}^{net}$, (e) net strangeness density 
$\rho_\mathrm{S}^{net}$, (b) temperature $T_\mathrm{SM} $, (d) baryon 
chemical potential $\mu_\mathrm{B}$, and (f) strangeness chemical 
potential $\mu_\mathrm{S}$ in the central cell with $V = 125$~fm$^3$ in
central Au+Au collision calculated within UrQMD at energies $E_{lab} = 
10A$~GeV (circles), $20A$~GeV (triangles), $30A$~GeV (squares), and 
$40A$~GeV (diamonds). Lines are drawn to guide the eye.}
        \label{fig:rho_t_mu}
\end{figure}

First, we study the time evolution of the bulk characteristics in 
central cell of Au+Au collisions at four energies in question. Entropy
density, net baryon density, and net strangeness density obtained in the 
cell from the microscopic calculations at time $1 \leq t \leq 20$~fm/$c$ 
are displayed in Figs.~\ref{fig:rho_t_mu}(a), \ref{fig:rho_t_mu}(c), and
\ref{fig:rho_t_mu}(e). At lowest bombarding energy $E_{lab} = 10A$~GeV 
the maximum values of $\varepsilon$ and $\rho_{\rm B}$ are reached at $t 
\approx 5$~fm/$c$, corresponding to complete overlap of two colliding 
nuclei. With rising bombarding energy the nuclei overlap occurs earlier, 
thus the maxima of the distributions are shifted to times $t \approx 
1-3$~fm/$c$. With the net strangeness in the cell the situation is more 
peculiar. Copious production of strange particles takes place between 
4~fm/$c$ and $8-10$~fm/$c$ when the matter in the cell is baryon rich. 
As mentioned in \cite{plb_98,jpg_99,prc_99,prc_08}, $K^+$'s can leave 
the selected volume a bit earlier compared to the $K^-$'s because of the 
smaller interaction cross sections. Therefore, the net strangeness in 
the cell is always negative, though small. Applying the procedure 
explained in Sec.~\ref{sec:method} we insert the values of 
$\{\varepsilon, \rho_{\rm B}, \rho_{\rm S} \}$ as an input in the SM to 
get $\{ T,\mu_{\rm B}, \mu_{\rm S} \}$ corresponding to an ideal hadron 
gas in chemical and thermal equilibrium. Evolutions of these parameters 
are shown in Figs.~\ref{fig:rho_t_mu}(b), \ref{fig:rho_t_mu}(d), and
\ref{fig:rho_t_mu}(f). It is worth noting that the local equilibrium in 
the cell at energies between $10A$~GeV and $40A$~GeV is reached not 
earlier than $t \approx 6-8$~fm/$c$. Therefore, one should treat the SM 
parameters obtained for earlier times with great care. 
Large baryon and energy densities observed at $t \leq 6$~fm/$c$ are
caused by interpenetration of two Lorentz-contracted nuclei. This leads
to extra-high temperatures of the ideal hadron gas, seen in
Fig.~\ref{fig:rho_t_mu}(b). For the extraction of more reliable values 
of $T$ and $\mu_{\rm B}$ we have to wait until the remnants of colliding 
nuclei will pass through each other and leave the tested volume. From
here, we will indicate the thermodynamic results related to the early 
phase of the matter evolution in the cell by dashed lines in the figures.

Despite the differences in the cell initial conditions, all four 
temperature curves sit on the top of each other after $t = 7$~fm/$c$.
Both baryon and strangeness chemical potentials drop with increasing
bombarding energy, in full accord with the SM analysis of experimental 
data. However, $\mu_{\rm B}$ increases whereas $\mu_{\rm S}$ decreases, 
while the temperature in the cell drops and the matter becomes more 
dilute.

\begin{figure}
\resizebox{\linewidth}{!}{
\includegraphics[scale=0.60]{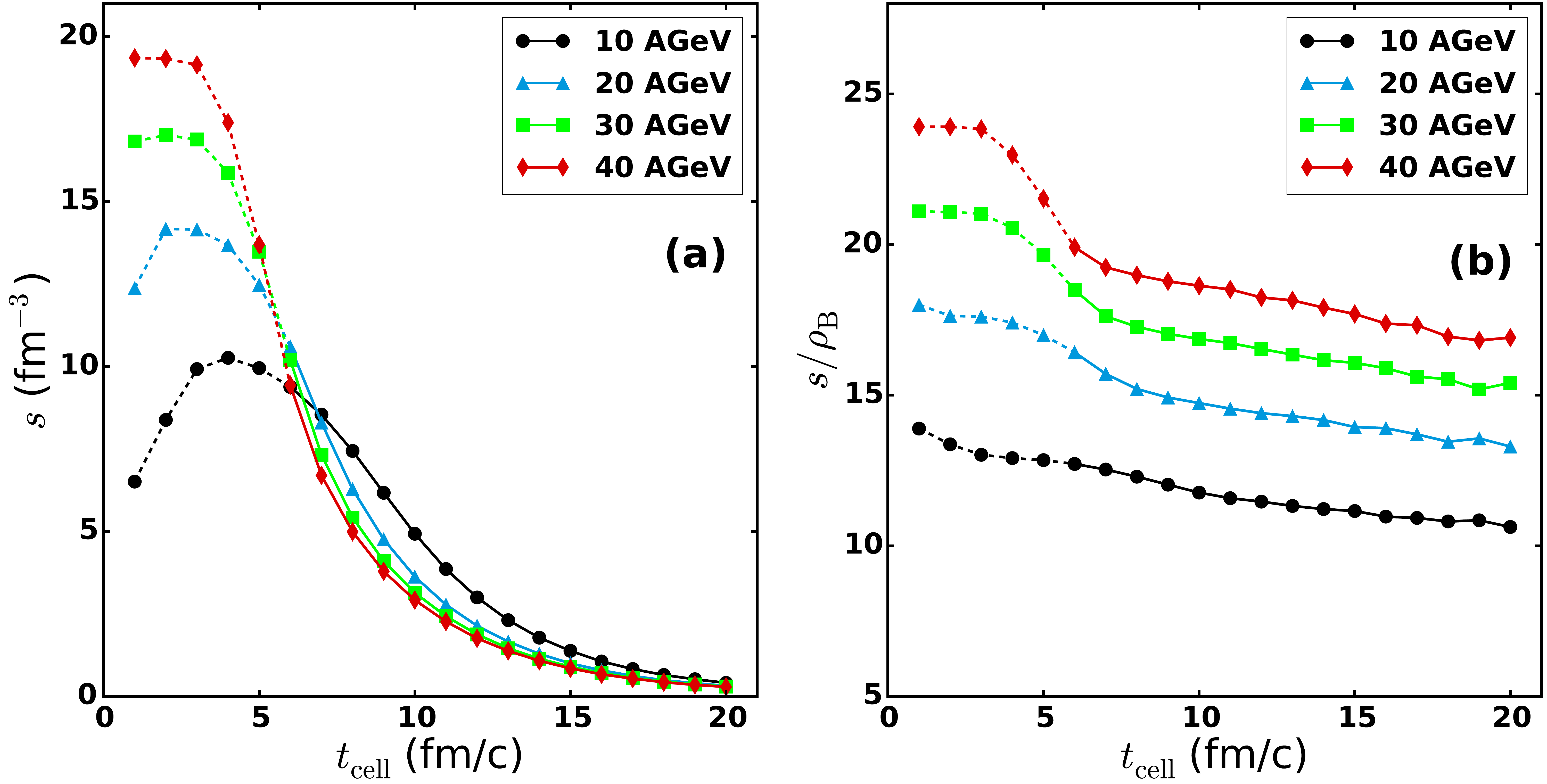}
}
\caption{(Color online)
(a) Entropy density $ s_\mathrm{sm}$ and (b) its ratio to net 
baryon density $ s_\mathrm{sm} / \rho_\mathrm{b, net} $ for different 
collision energies $E$ in UrQMD central cell calculations. Lines are 
drawn to guide the eye.}
\label{fig:s_sm_tcell}
\end{figure}

Figure~\ref{fig:s_sm_tcell}(a) presents the evolution of the entropy 
density in the central cell in the studied reactions. This behavior is
qualitatively similar to that of $\varepsilon(t)$ seen in 
Fig.~\ref{fig:rho_t_mu}(a). Note, however, that the entropy density here
is calculated within the SM implying the maximum values for $s$.  For 
the nonequilibrium state at $t \leq 6$~fm/$c$ the entropy density is 
lower than the $s^{SM}$. The ratio of entropy density to baryon density,
$s / \rho_{\rm B}$, shown in Fig.~\ref{fig:s_sm_tcell}(b), also should 
be lower during the stage of relaxation to equilibrium. It drops 
slightly about 15\% between 6 and 20~fm/$c$ indicating that the
expansion proceeds nearly isentropically.

\begin{figure}
	\includegraphics[scale=0.13]{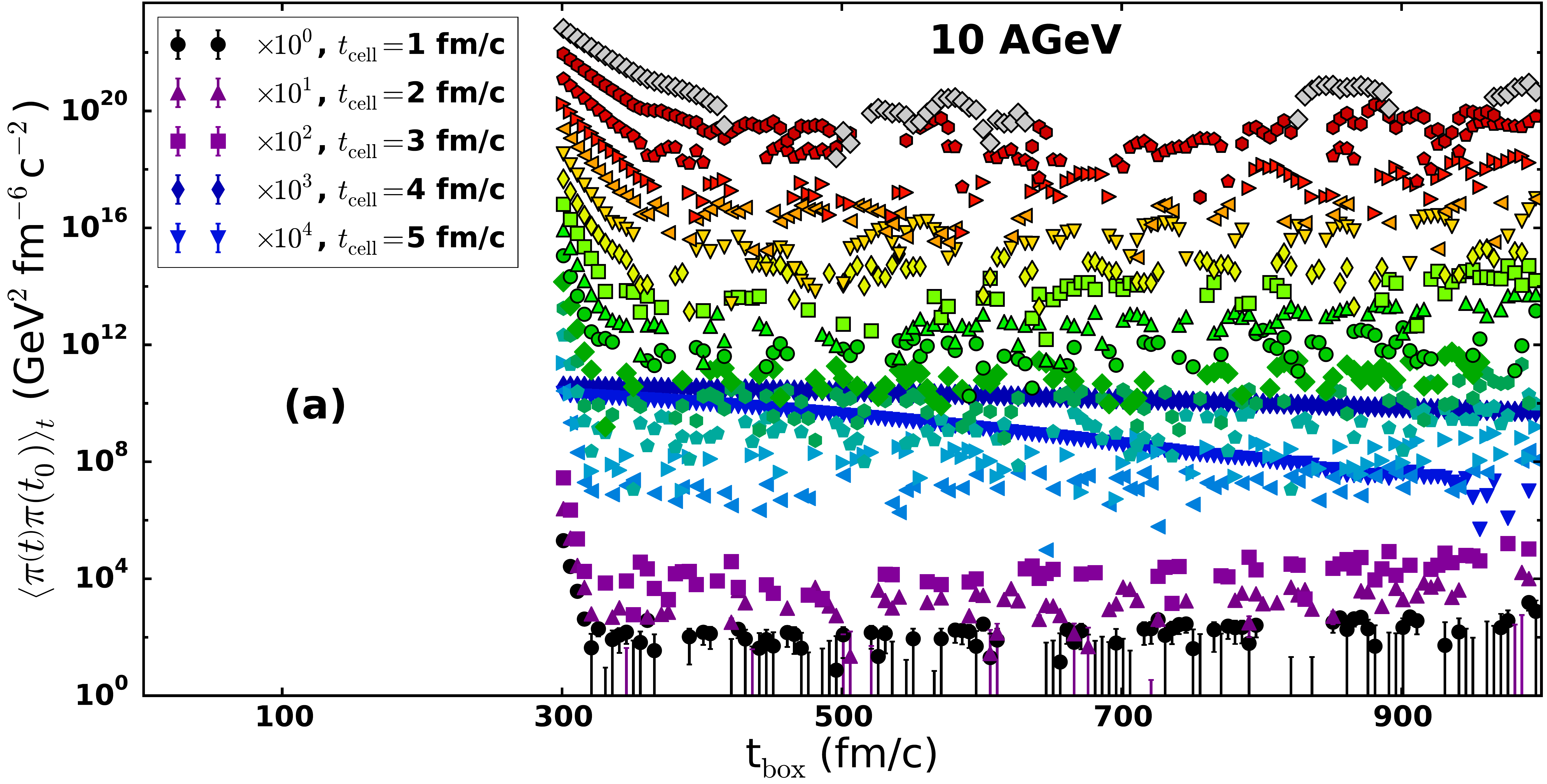}
	\includegraphics[scale=0.13]{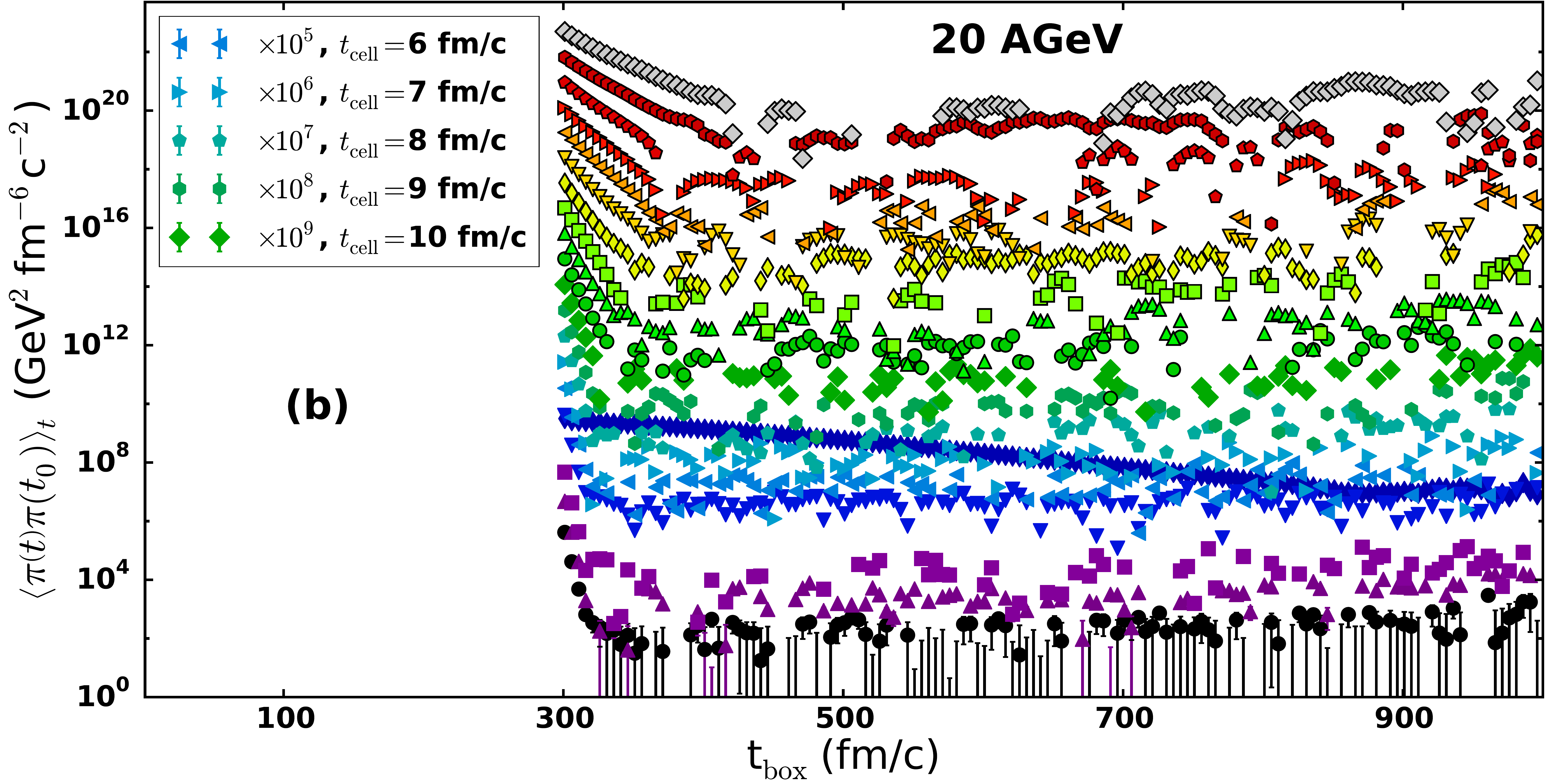}
	\includegraphics[scale=0.13]{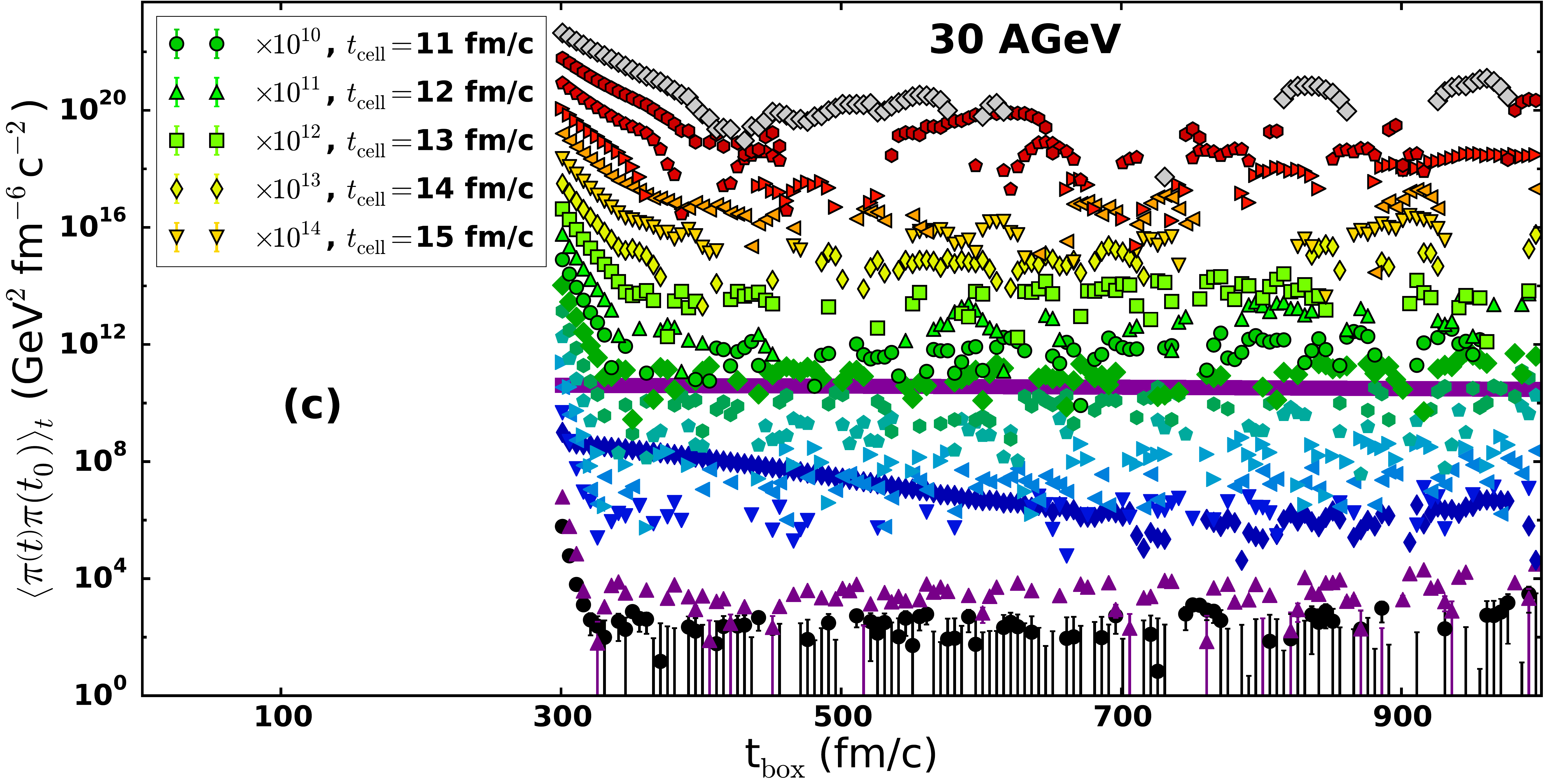}
	\includegraphics[scale=0.13]{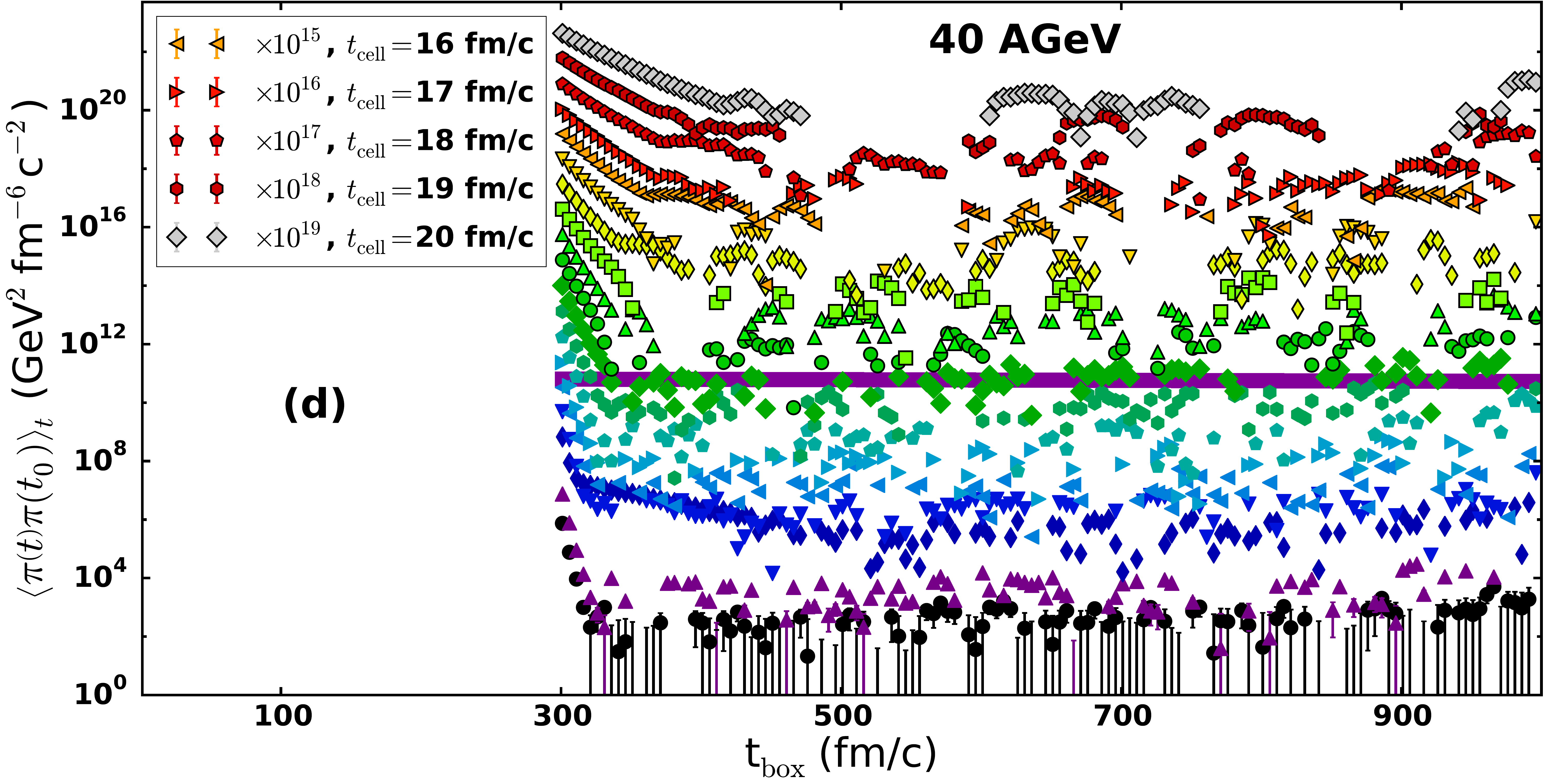}
\caption{(Color online)
Correlators $ \langle \pi\left(t\right) \pi\left(t_0\right) 
\rangle_t $ for initial cutoff time $t_0 = 300$~fm/$c$ in the UrQMD 
box calculations. Initial conditions for the boxes are taken from the
central cell with $V = 125$~fm$^3$ of Au+Au collisions at (a) $E_{lab} =
10A$~GeV, (b) $20A$~GeV, (c) $30A$~GeV, and (d) $40A$~GeV at times $t = 
1 - 20$~fm/$c$. Each distribution is multiplied by factor 
$10^{t_{cell}-1}$.}
\label{fig:correlator_300}
\end{figure}

We are switching now to the box calculations. 
Figure~\ref{fig:correlator_300} shows correlators defined by 
Eq.~(\ref{correlator}) calculated for all four collision energies. The
input data $\varepsilon, \rho_{\rm B}, \rho_{\rm S}$ were extracted from
the central cell of Au+Au central collisions at times from 1~fm/$c$ up 
to 20~fm/$c$ after the beginning of the collision. To see the 
differences between the distributions more distinctly, each correlator 
was multiplied by the factor $10^{t_{cell} - 1}$. Recall, that the 
results of the box calculations are shown for times $t_{box} \geq 
300$~fm/$c$. This timescale has nothing to do with the typical 
relaxation times of hot and dense matter in heavy-ion collisions 
\cite{prc_00}. One can see that all correlators reveal exponential 
falloff with time in accordance with Eq.~(\ref{correlator_exp}). However, 
for the conditions corresponding to early cell times, the relaxation 
rates are several orders of magnitude slower compared to those 
corresponding to late times. This cannot be explained entirely by large 
baryon and energy densities in the central cell at early 
$t_\mathrm{cell}$, when nuclei overlap. Here one has to initialize the 
box with one or two very ultrarelativistic kaons that cannot 
redistribute their energy and momenta quickly enough. This circumstance 
results in a slow relaxation of the appropriate correlators. In order to 
extract the correct data corresponding to the overlap of nuclei one has 
to process the box calculations for longer periods of time; see, e.g., 
\cite{PRC.97.055204}. Note also that microscopic transport models 
usually lack the inverse reactions to multiparticle processes $2 
\rightarrow N,\ (N \geq 3)$. In this case the matter in the box will 
relax to the steady state rather than to the pure equilibrium; see, 
e.g., \cite{prc_98,prc_00,npa_99,HSD_00}. However, the
matter in the central cell at $t \geq 6$~fm/$c$ in heavy-ion collisions
at energies below $E_{lab} = 40A$~GeV becomes dilute very quickly. Its
energy density drops, and the many-particle inelastic reactions in the
box with similar $\varepsilon, \rho_{\rm B}$, and $\rho_{\rm S}$ 
rapidly cease, thus leading to equilibrium similar to that of the SM.

At late times of the box calculations it appears that the correlations 
are rising. This is a technical effect. Namely, at the end of the UrQMD 
box calculations the program forces decay of all strongly decaying 
resonances, which may lead to some momentum correlations.

\begin{figure}
\resizebox{\linewidth}{!}{
	\includegraphics[scale=0.60]{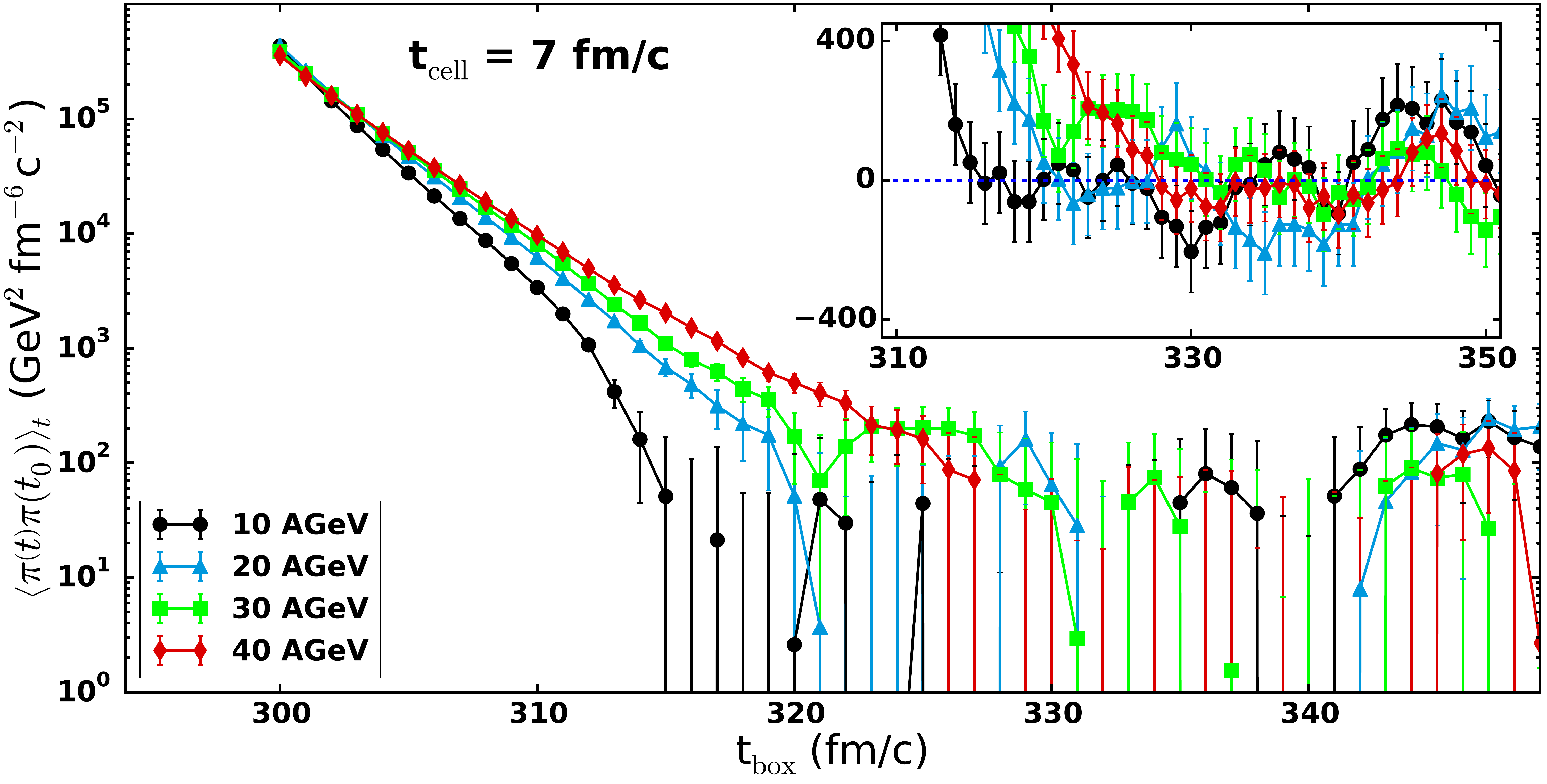}
}
\caption{(Color online)
UrQMD box calculations of the correlators 
$\langle \pi\left(t\right) \pi\left(t_0\right) \rangle_t$. Initial
conditions in the box correspond to those in the central cell of Au+Au
collisions at $E_{lab} = 10A$~GeV (circles), $20A$~GeV (triangles),
$30A$~GeV (squares), and $40A$~GeV (diamonds) taken at time 
$t_\mathrm{cell} = 7$~fm/$c$. Lines are drawn to guide the eye.}
\label{fig:correlator_300_7}
\end{figure}

Typical behavior of the correlator dynamics on shorter timescales is
demonstrated in Fig.~\ref{fig:correlator_300_7}, where the correlators 
for different collision energies are depicted. Again, as in 
Fig.~\ref{fig:correlator_300}, the initial cutoff time in the box is
$t_0 = 300$~fm/$c$. The initial conditions in the box correspond to that
in the cell at $t_{cell} = 7$~fm/$c$. The exponential falloff with time 
occurs within $t \lesssim t_0 + 30$~fm/$c$. After that time the 
correlators become too weak, and fluctuations start to dominate the 
system. Domination of the fluctuations leads to the necessity of 
cutting off the dataset while fitting the correlator 
$\langle \pi\left(t\right) \pi\left(t_0\right) \rangle_t $ to 
Eq.~(\ref{correlator_exp}), as was proposed in 
\cite{PRC.86.054902,PRC.97.055204}.

The necessity for dataset cutoff raises up the question of direct 
applicability of Eq.~(\ref{eta}) in numerical calculations. In order to 
investigate the problem, we compare next the relaxation times extracted 
both from the integral in Eq.~(\ref{eta}), $\tau_\mathrm{int}
\left(t_0\right)$, and by fitting the correlator to 
Eq.~(\ref{correlator_exp}) within the time interval cutoff 
$t_0 \leq t \leq \left(t_0 + 30\right)$~fm/$c$, 
$\tau_\mathrm{fit}\left(t_0\right)$.

\begin{figure}
\resizebox{\linewidth}{!}{
	\includegraphics[scale=0.60]{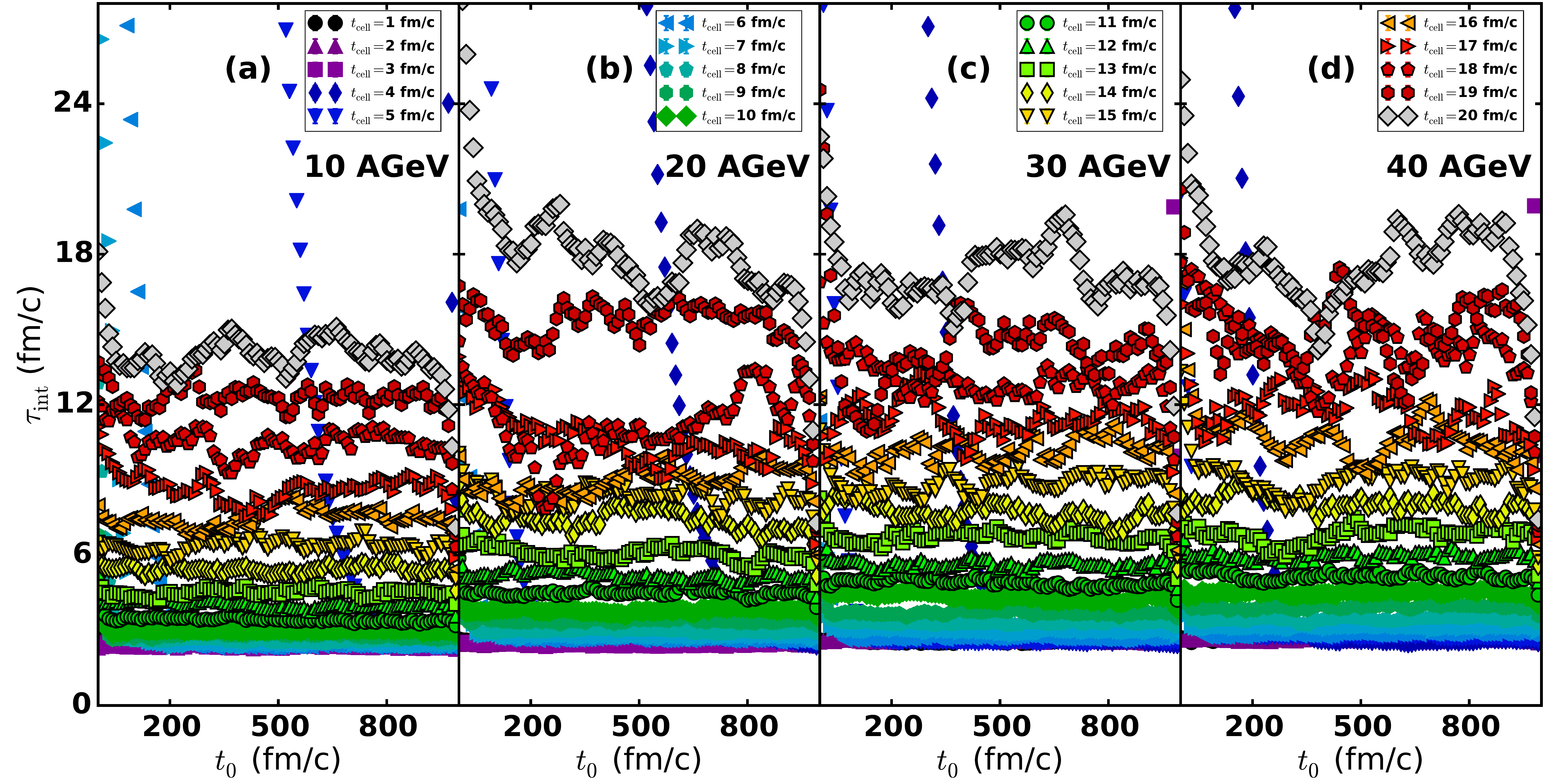}
}
\caption{(Color online)
Relaxation time $\tau_\mathrm{int}\left(t_0\right)$ for the collision 
energies (a) $E_{lab} = 10A$~GeV, (b) $20A$~GeV, (c) $30A$~GeV, and (d)
$40A$~GeV and for all cell times $1 \leq t_\mathrm{cell} \leq 20$~fm/$c$ 
in the UrQMD box calculations.}
\label{fig:tau_int_t0}
\end{figure}

Figure~\ref{fig:tau_int_t0} depicts the dependence of relaxation time 
$\tau_\mathrm{int}$, extracted from the integral in Eq.~(\ref{eta}), on 
the initial cutoff time $t_0$, with every tenth point being shown. As 
one can see, the relaxation usually takes a longer period for $t_0$ 
shorter than 200~fm/$c$ and vanishes for $t_0 \geq 900$~fm/$c$. For
the initial times between these two limits the relaxation time is rather
constant. The only exceptions are at the early cell times.

\begin{figure}
\resizebox{\linewidth}{!}{
	\includegraphics[scale=0.60]{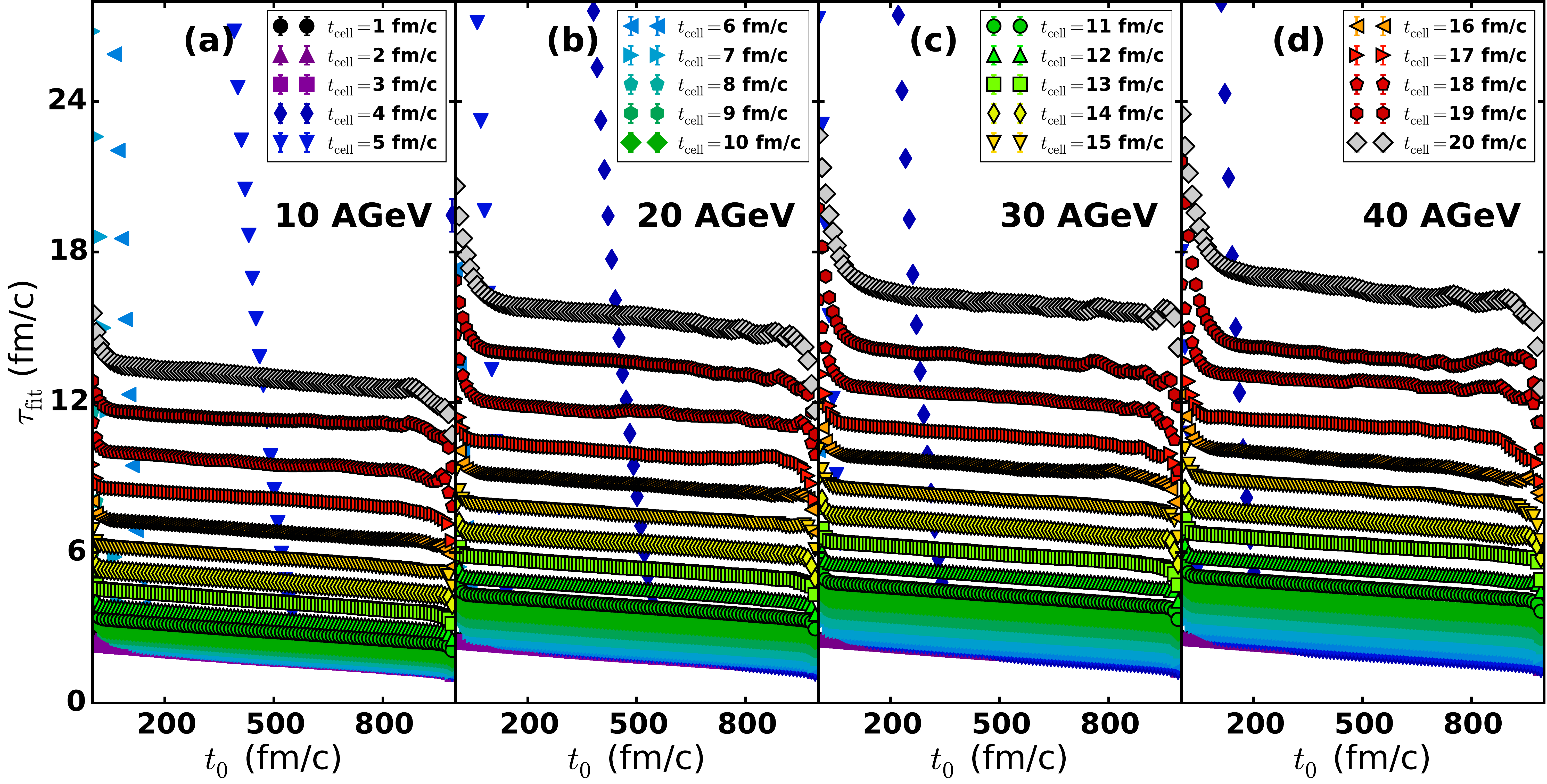}
}
\caption{(Color online)
The same as Fig.~\ref{fig:tau_int_t0} but for relaxation time 
$\tau_\mathrm{fit}\left(t_0\right)$.}
\label{fig:tau_fit_t0}
\end{figure}

Figure~\ref{fig:tau_fit_t0} displays the dependence of relaxation time 
$\tau_\mathrm{fit}$ extracted by fitting over the time interval 
$t_\mathrm{box}\in \left[t_0, t_0 + 30\right] $. The behavior of 
$\tau_\mathrm{fit} $ is pretty similar to that of $ \tau_\mathrm{int}$. 
However, the results presented in Fig.~\ref{fig:tau_fit_t0} have no 
stochastic oscillations, in contrast to those shown in 
Fig.~\ref{fig:tau_int_t0}. This can be explained by the influence of 
fluctuations on $\tau_\mathrm{int}$. It is worth mentioning that, as one 
can notice, the plateau demonstrates some slope in 
Fig.~\ref{fig:tau_fit_t0} at $t_0 \geq 200$~fm/$c$ as compared to the 
results shown in Fig.~\ref{fig:tau_int_t0}. The slope may significantly 
influence the determination of $\eta$ values, because for early cell 
times with minimum values of $\tau_\mathrm{fit}$ it may vary 
approximately by 40\% for $200 \leq t_0 \leq 800$~fm/$c$.
Small values of $\tau_\mathrm{int (fit)}$ at large $t_0$ are dealing 
with the small averaging interval; see Eq.~(\ref{correlator}). Namely, 
the time resolution at large $t_0$ is too high to observe the correlator 
falloff, and one finds a kind of Brownian motion instead.

For the midrange of the initial cutoff time $t_0$ at the plateau $-$ see 
Figs.~\ref{fig:tau_int_t0} and \ref{fig:tau_fit_t0} $-$ the falloff 
rate does not change significantly. Thus, the values of $t_0$ from 
this range are well suited for our task. In the following we average the 
value of $\tau_\mathrm{int (fit)}$ over the plateau in order to reduce 
statistical errors. Large values of the relaxation time 
$\tau_\mathrm{int/fit}$ for some early cell times $t_\mathrm{cell}$ are 
explained by the copious production of new hadrons and their subsequent 
rescatterings in very hot and dense baryon-rich matter at the very 
beginning of the collision. Additional time delay is caused by energetic 
single negative kaons. Combination of these factors forces the extension 
of the box calculations up to 2000 (sometimes 3000) fm/$c$. 

\begin{figure}
\resizebox{\linewidth}{!}{
        \includegraphics[scale=0.60]{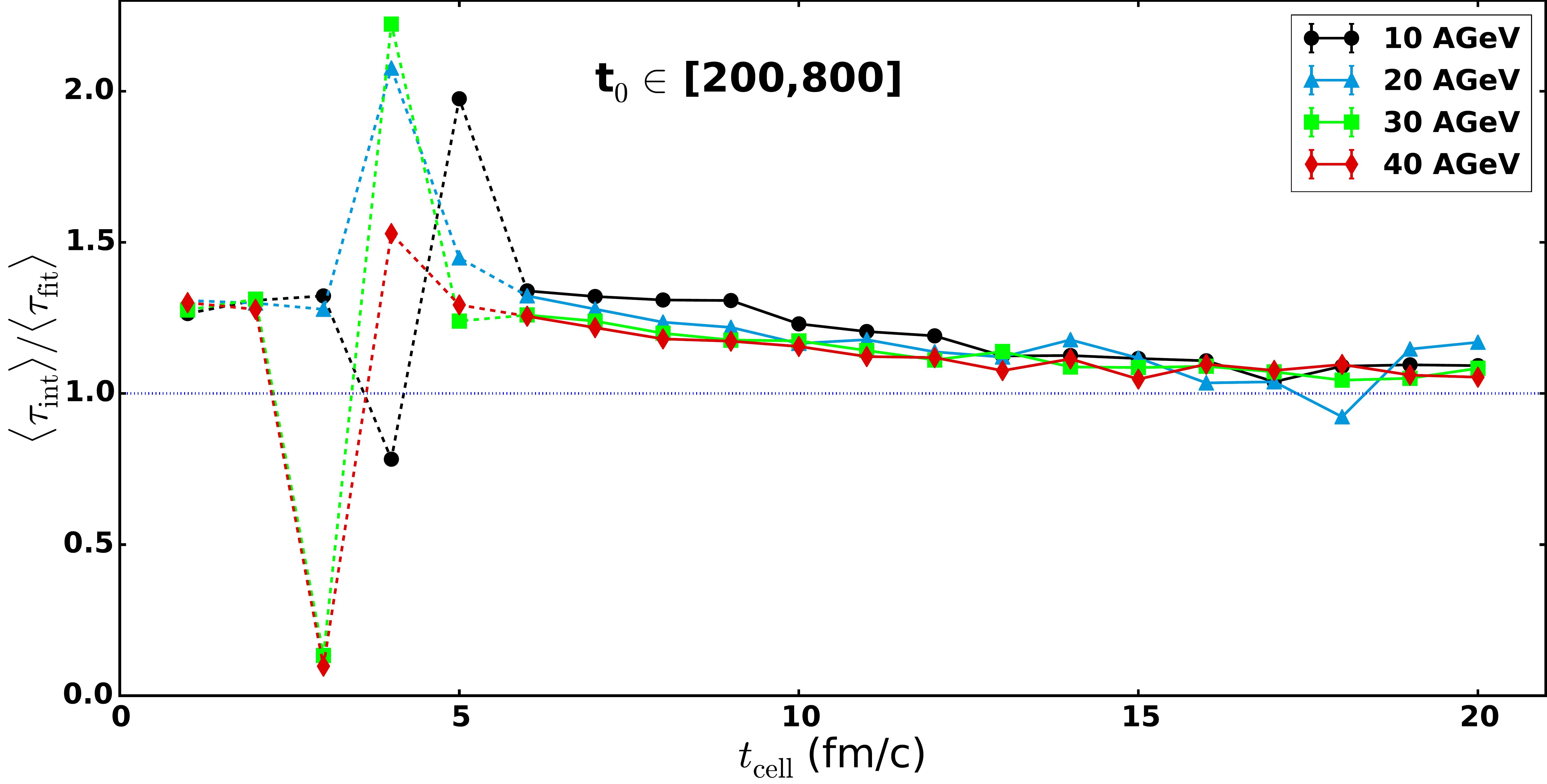}
}
\caption{(Color online)
Ratio 
$\langle\tau_\mathrm{int}\rangle/\langle\tau_\mathrm{fit}\rangle$ 
for the collision energies $10A$~GeV (circles), $20A$~GeV (triangles),
$30A$~GeV (squares), and $40A$~GeV (diamonds) for all cell times 
$t_\mathrm{cell}$. Errors are smaller than the symbol sizes.}
\label{fig:tau_comparison_tcell}
\end{figure}

Figure~\ref{fig:tau_comparison_tcell} shows ratio of the relaxation 
times determined by Eqs.~(\ref{eta}) and (\ref{correlator_exp}),
$\langle\tau_\mathrm{int}\rangle/\langle\tau_\mathrm{fit}\rangle$. 
As we see, $\tau_\mathrm{int}$ exceeds $\tau_\mathrm{fit}$ by 
$25\%$ at $t = 6$~fm/$c$. For the cell conditions at later stages the
relaxation times converge and agree with each other within 10\%
accuracy at $t \geq 15$~fm/$c$. Thus, taking the fluctuations into 
account results in increase of $\tau$, as well as in its noise-like 
oscillations. The only difference, except for the general slope of 
$\tau_\mathrm{fit}$, is observed at the early cell times, when the 
nuclei overlap.

\begin{figure}
\resizebox{\linewidth}{!}{
	\includegraphics[scale=0.60]{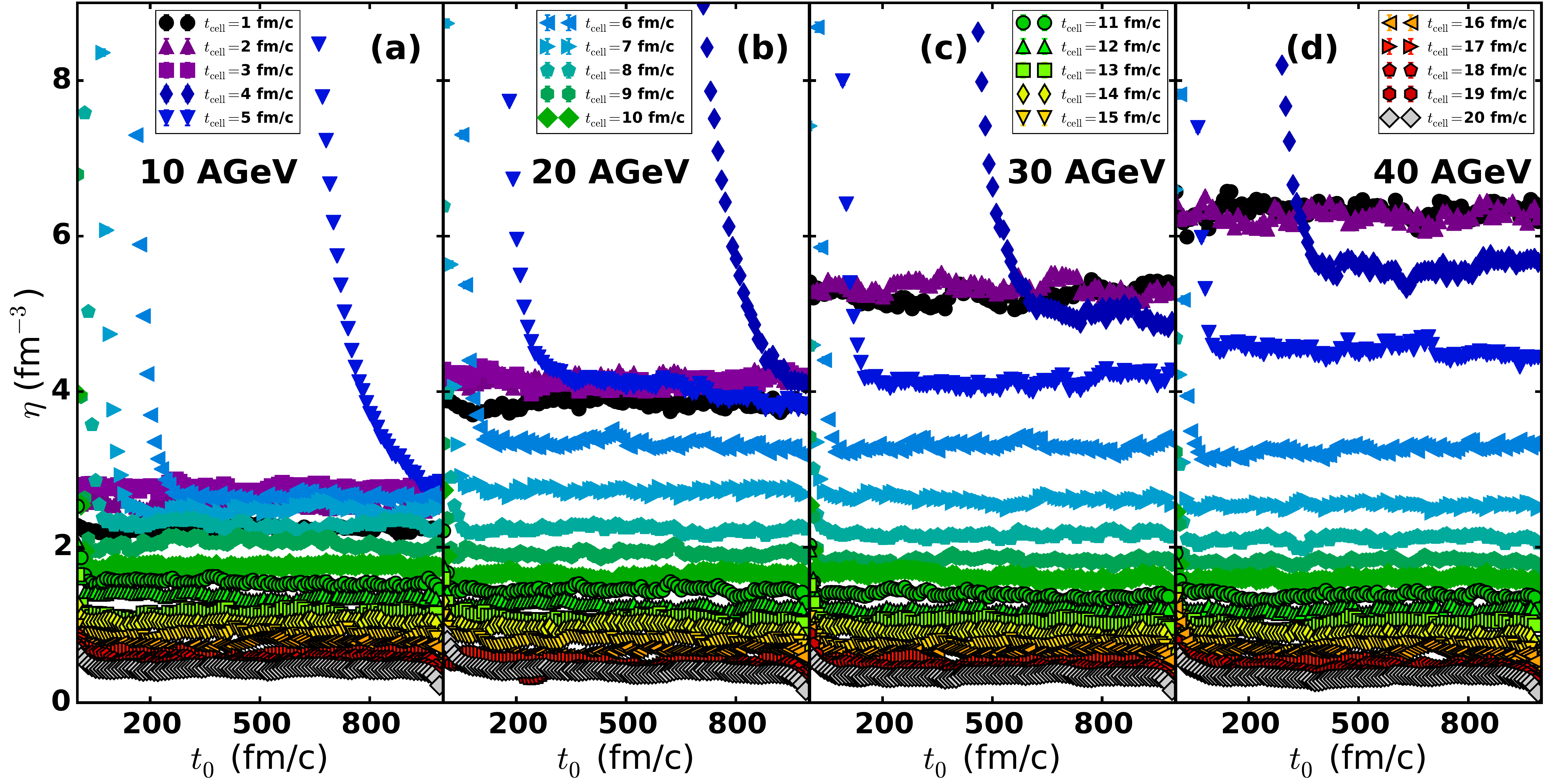}
}
\caption{(Color online)
Shear viscosity $\eta\left(t_0\right)$ for the collision 
energies (a) $E_{lab} = 10A$~GeV, (b) $20A$~GeV, (c) $30A$~GeV, and (d) 
$40A$~GeV for all cell times $1 \leq t_\mathrm{cell} \leq 20$~fm/$c$
within the UrQMD box calculations.}
\label{fig:eta_t0}
\end{figure}

Shear viscosity $\eta\left(t_0\right)$, calculated with 
$\tau_\mathrm{int}$, is presented in Fig.~\ref{fig:eta_t0}. Since $\eta$ 
is proportional to $\tau_\mathrm{int}$ due to exponential falloff 
behavior of the correlator, distributions in Figs.~\ref{fig:tau_int_t0} 
and \ref{fig:eta_t0} have many similar features. Shear viscosity shows 
larger values for the initial box fluctuations at small times $t_0$. It 
is reduced significantly at large $t_0$, and has a plateau at 
intermediate times.

\begin{figure}
\resizebox{\linewidth}{!}{
	\includegraphics[scale=0.60]{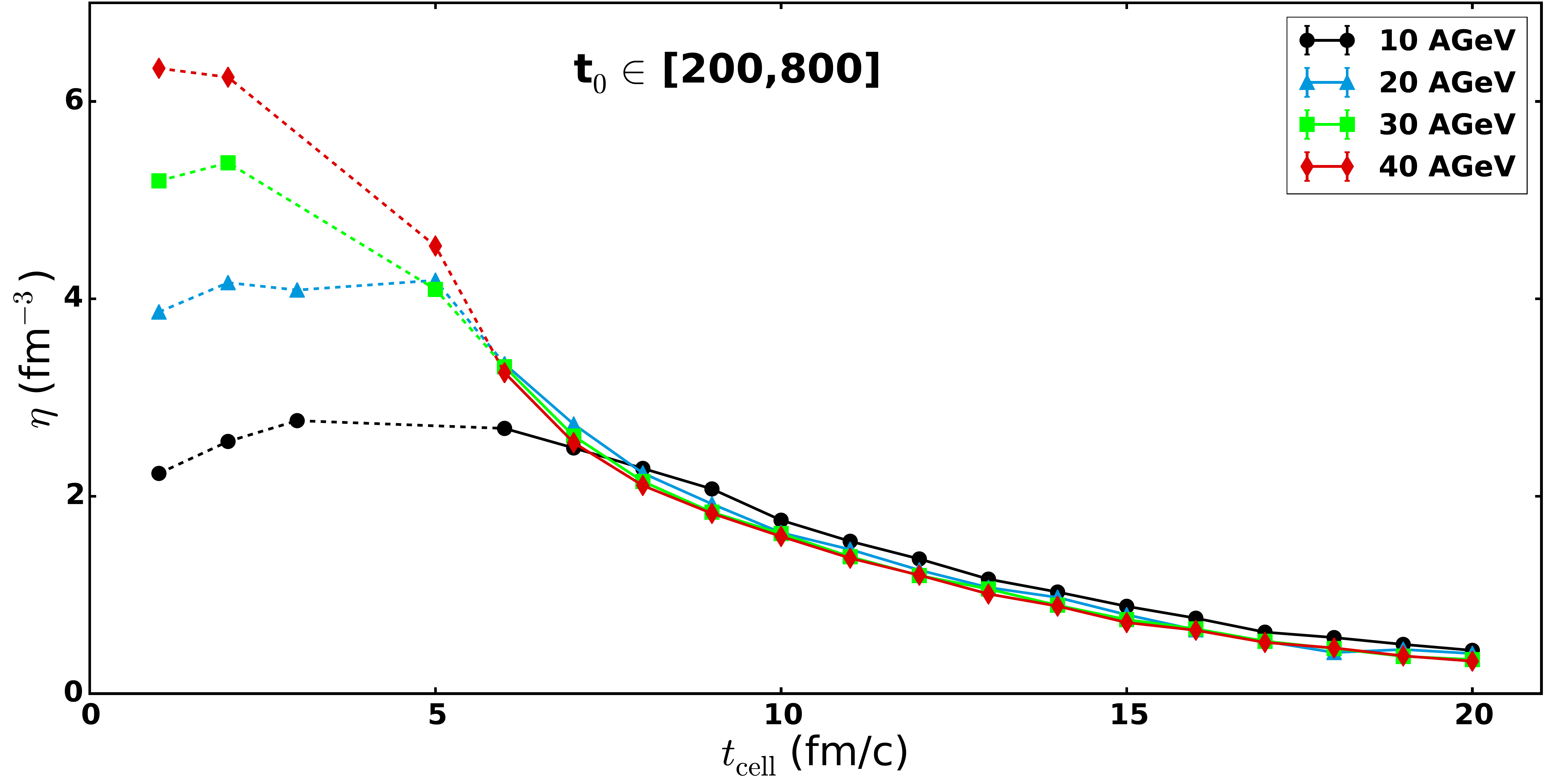}
}
\caption{(Color online)
Shear viscosity $\eta\left(t_\mathrm{cell}\right)$ of hadrons
in the central cell of central Au+Au collisions at (a) $E_{lab} = 
10A$~GeV, (b) $20A$~GeV, (c) $30A$~GeV, and (d) $40A$~GeV within the 
UrQMD box calculations. Lines are drawn to guide the eye.}
\label{fig:eta_tcell}
\end{figure}

After averaging over the plateau, which we define as $t_0 \in \left[200, 
800\right]$~fm/$c$, one may obtain shear viscosity for different cell 
times at all the collision energies considered. Results are shown in
Fig.~\ref{fig:eta_tcell}. The statistical errors are smaller than the 
symbol sizes. We see that shear viscosity reaches its maximum at the 
very beginning of the heavy-ion collision. Then it gradually drops 
almost to zero at the late cell times. Decrease of $\eta$ with time is 
explained by the fact that at the late stages of the evolution of 
nuclear matter in the central cell there are only (quasi)elastic 
processes, i.e., soft scattering modes, remaining \cite{prc_99}. All 
energetic hadrons with large momenta have already left the cell. This 
circumstance results in the fast redistribution of momentum and energy 
of soft hadrons over the system, and, consequently, in small relaxation 
rate $\tau$ of the correlator.

At early times the shear viscosity is larger for heavy-ion collisions at
larger energies. But after $t \approx 6$~fm/$c$ all curves representing 
four different energies quickly converge. This behavior is very similar 
to the drop of the cell temperatures shown in 
Fig.~\ref{fig:rho_t_mu}(d). Both effects are caused by the faster loss
of energy and baryon density in the central cell of central collisions
with increasing bombarding energies.   

\begin{figure}
  \includegraphics[scale=0.13]{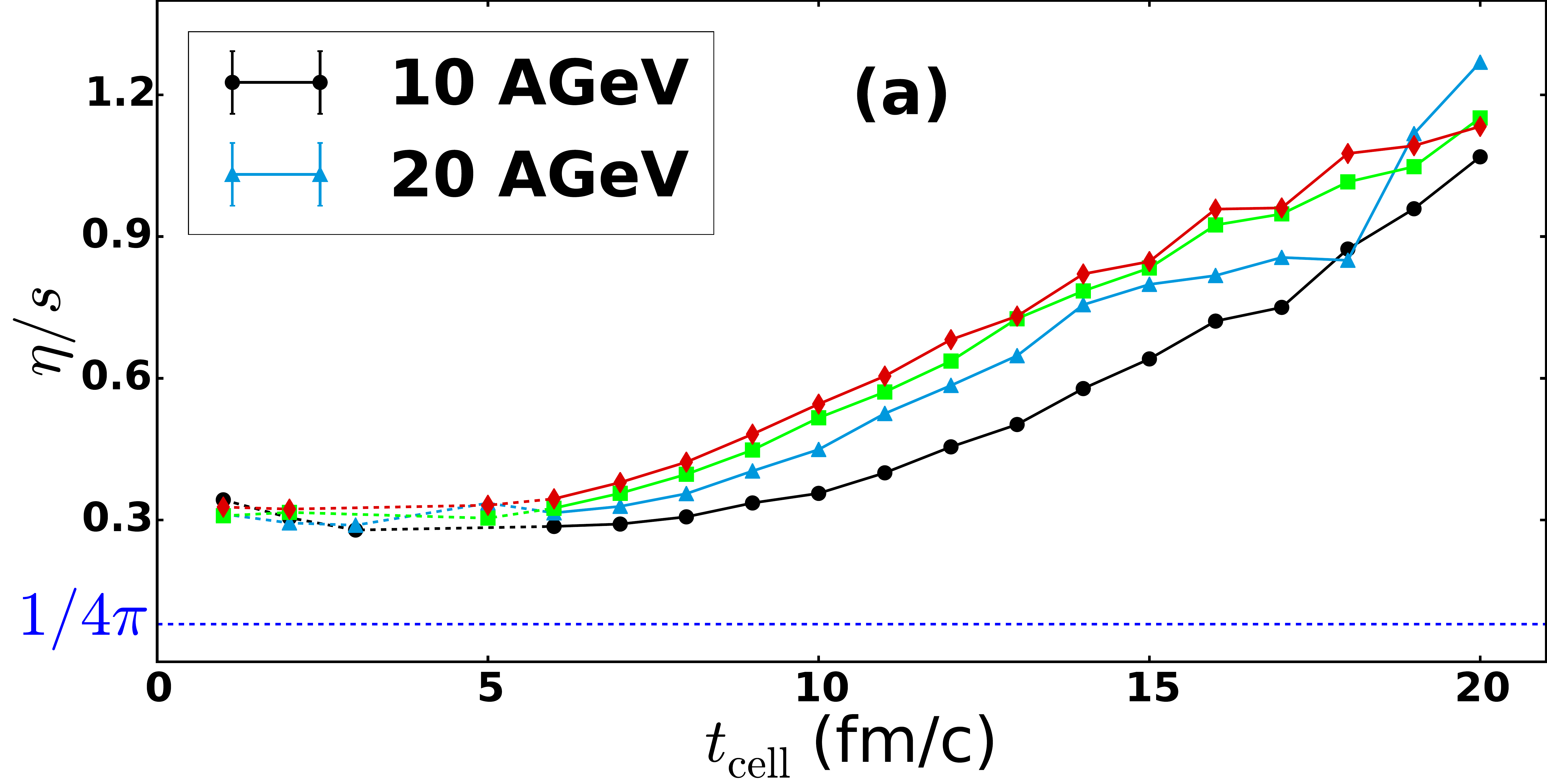}
  \includegraphics[scale=0.13]{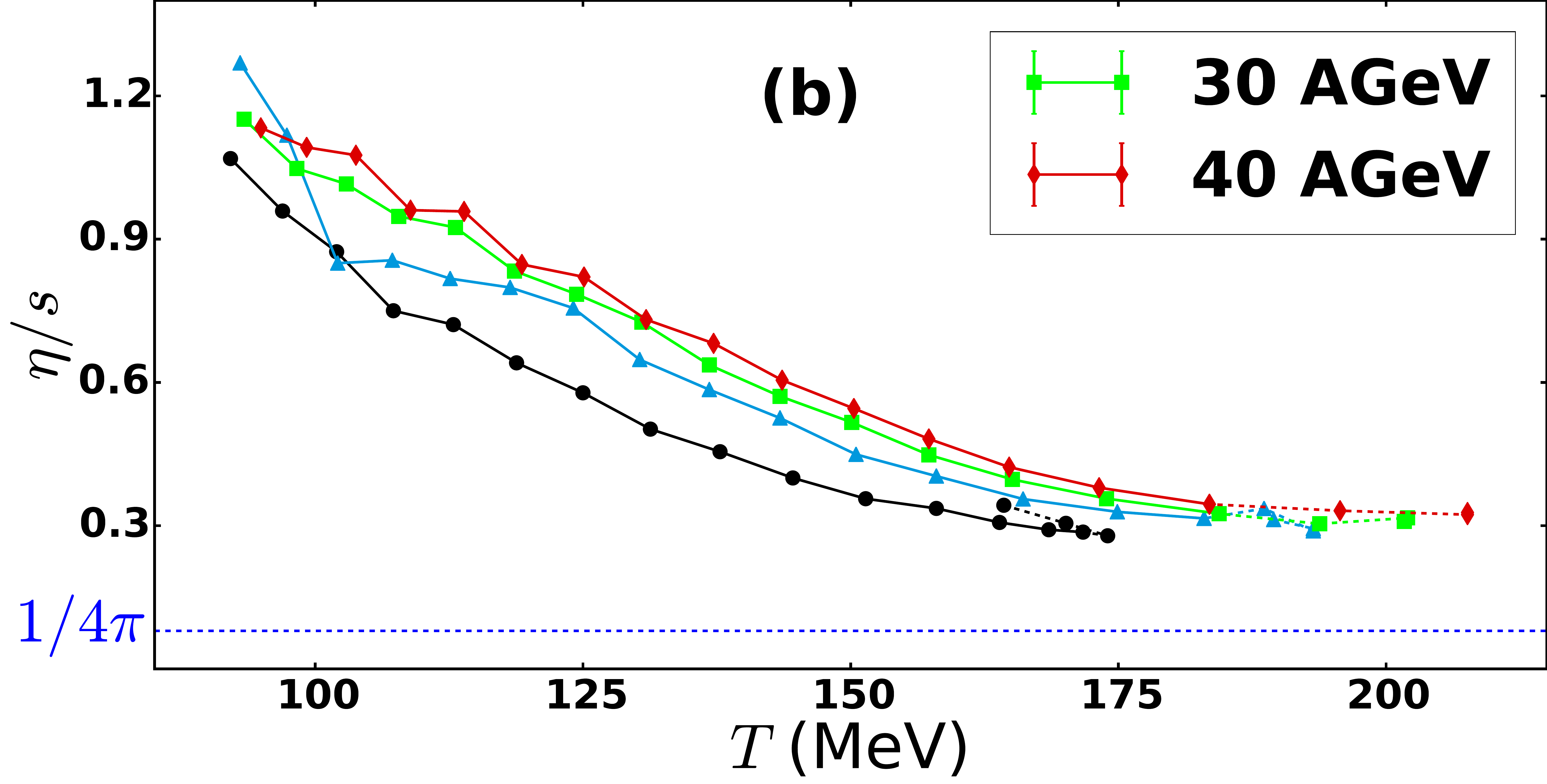}
  \includegraphics[scale=0.13]{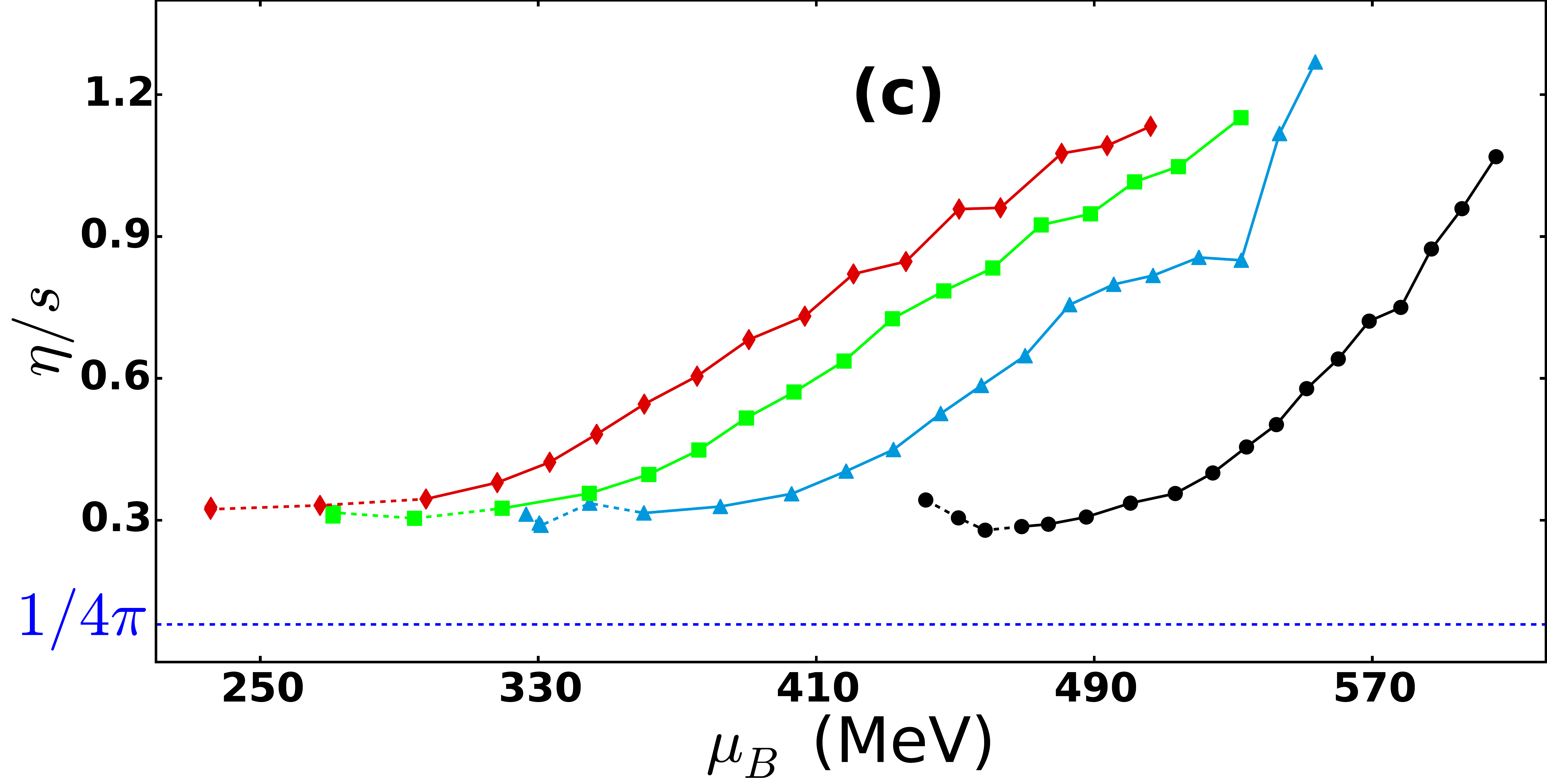}
  \includegraphics[scale=0.13]{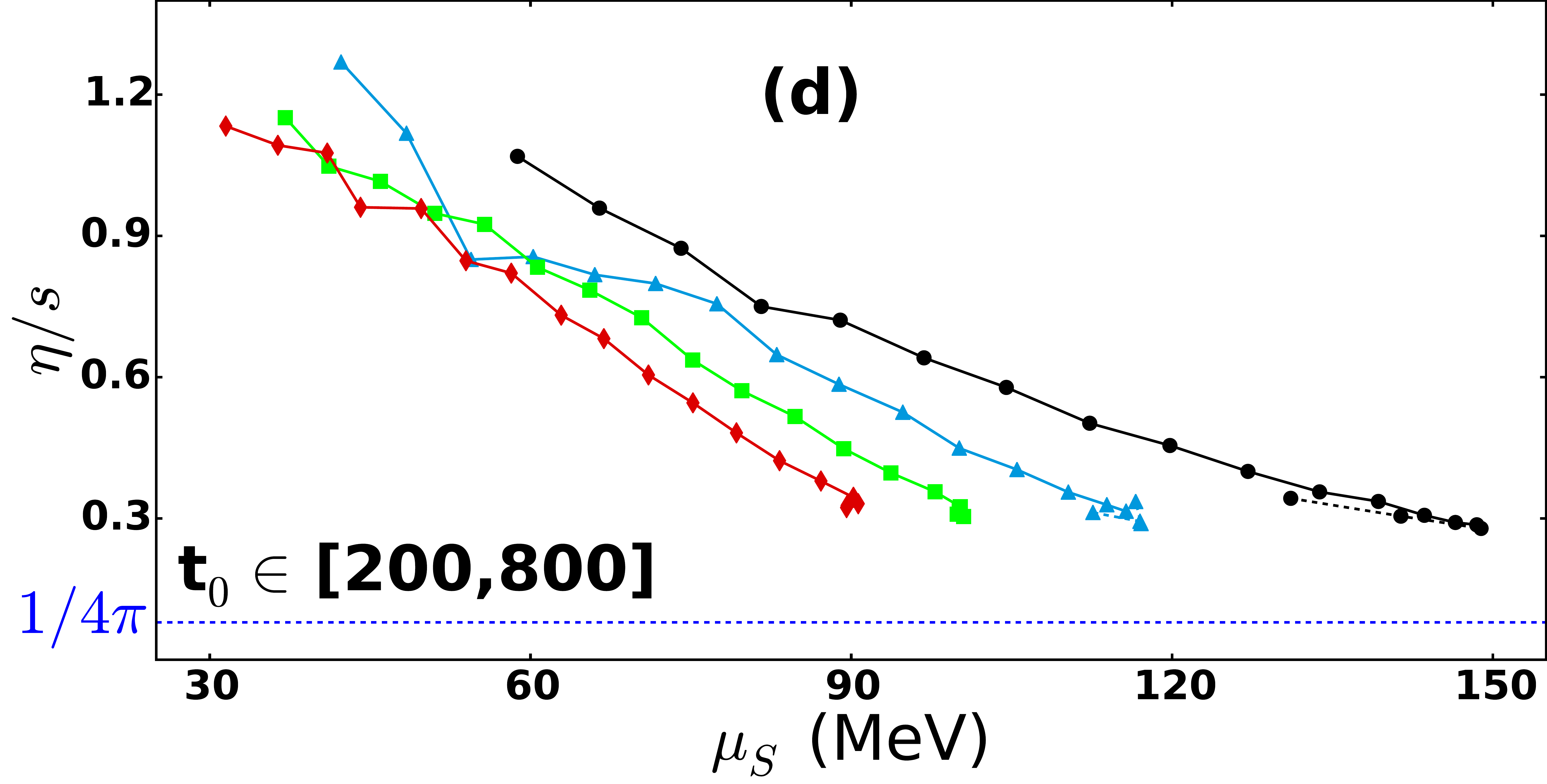}
\caption{(Color online)
Shear viscosity to SM entropy ratio $\eta / s_\mathrm{sm}$ as function
of (a) time (b) temperature, (c) baryon chemical potential, and (d)
strangeness chemical potential in the UrQMD calculations of central cell 
of central Au+Au collisions at $E_{lab} = 10A$~GeV (circles), $20A$~GeV 
(triangles), $30A$~GeV (squares), and $40A$~GeV (diamonds). Lines are 
drawn to guide the eye.}
\label{fig:eta_to_s_sm}
\end{figure}

Finally, Fig.~\ref{fig:eta_to_s_sm} displays $\eta / s$ dependencies on 
the evolution of the cell parameters, i.e., time 
[Fig.~\ref{fig:eta_to_s_sm}(a)], SM temperature 
[Fig.~\ref{fig:eta_to_s_sm}(b)], baryon chemical potential
[Fig.~\ref{fig:eta_to_s_sm}(c)], and strangeness chemical potential
[Fig.~\ref{fig:eta_to_s_sm}(d)]. The statistical errors are smaller than 
the symbol sizes. For all energies the ratio $\eta / s$ reaches its 
minimum at $t \approx 5$~fm/$c$, when the nuclei are expected to 
overlap. Despite of being small enough, the minima are about four times 
larger than the theoretical minimum value $1/4\pi$. After that the ratio 
$\eta / s$ in the cell increases with time. The lower the collision 
energy, the smaller the ratio. It is also increasing with the drop of
temperature and strangeness chemical potential, as shown in 
Figs.~\ref{fig:eta_to_s_sm}(b) and \ref{fig:eta_to_s_sm}(d), and
with the rise of baryochemical potential; see 
Fig.~\ref{fig:eta_to_s_sm}(c). It is worth noting that at $t \leq 
5$~fm/$c$ the matter in the cell is still out of equilibrium, whereas
the estimates of $T, \mu_{\rm B}$, and $\mu_{\rm S}$ are done for a 
fully equilibrated system of hadrons. Therefore, all distributions at 
early times are indicated by the dashed curves.     

Comparing our results to those calculated within the SMASH model in
\cite{PRC.97.055204}, one can notice a qualitatively different 
dependence of $\eta / s^\mathrm{SM}$ on the temperature. In contrast to 
the rise of $\eta / s$ with the temperature drop in the UrQMD cell
calculations, SMASH demonstrates almost constant behavior of this ratio
within the same temperature range. However, in the latter case the 
calculations were performed for a fixed baryon chemical potential, 
whereas in the UrQMD calculations it increases with the cell time 
$t_\mathrm{cell}$. Another reason for deviations is the nonzero 
strangeness chemical potential in our calculations. Nevertheless, as 
shown in \cite{PRC.97.055204}, the ratio $\eta / s$ increases in SMASH 
calculations with rise of baryon chemical potential, in accord with our 
results. Both UrQMD and SMASH indicate that shear viscosity decreases 
with decreasing temperature. This agreement is not accidental because of 
the conceptual similarity between UrQMD and SMASH. Further analysis 
concerning the influence of details of system's internal dynamics, 
particularly, the role of lifetimes of resonances, on the $\eta / s$ 
ratio can be found in \cite{PRC.97.055204}.

\section{Conclusions}
\label{sec:conclusions}

We have studied the shear viscosity of highly excited nuclear matter 
produced in the central area of central Au+Au collisions at energies 
$E_{lab} = 10A$, $20A$, $30A$, and $40A$~GeV. Calculations are done 
within the UrQMD model. At the first stage, the energy density, the net 
baryon density, and the net strangeness density are determined for a 
cubic central cell with volume $V = 125$~fm$^3$. After that, the 
obtained values are used as input to the statistical model of an ideal 
hadron gas to calculate temperature, baryon chemical potential, and 
strangeness chemical potential, as well as entropy density. The 
extracted values of $\varepsilon, \rho_{\rm B}$, and $\rho_{\rm S}$ are 
used also for initialization of the UrQMD box with periodic boundary 
conditions to study the relaxation of hot and dense nuclear matter to 
equilibrium. The Green-Kubo formalism is explored to calculate the shear 
viscosity. 

It is shown that equilibrium in the box is achieved approximately after 
$t  \geq 200$~fm/$c$ for all but very high baryon and energy densities,
corresponding to the overlap of the nuclei. The influence of initial 
cutoff time $t_0$ on momentum correlators is studied. Finally, the 
shear viscosity and its ratio to entropy density are calculated. We 
found that, for all four tested energies, $\eta$ and $s$ in the cell 
drop with time. Their ratios $\eta/s$, however, reach minima about 0.3 
at $t \approx5$~fm/$c$, irrespective of the bombarding energy. Then the
ratios rise to $\eta / s = 1.0 - 1.2$ at $t = 20$~fm/$c$. This increase 
is accompanied by the simultaneous rise of baryon chemical potential and 
drop of both temperature and strangeness chemical potential in the cell.
 
\begin{acknowledgments}
Fruitful discussions with K.~Bugaev, Yu.~Ivanov, D.~Olynichenko, and 
O.~Teryaev are gratefully acknowledged.
The work of L.B. and E.Z. was supported by Russian Foundation for Basic 
Research (RFBR) under Grants No. 18-02-40084 and No. 18-02-40085,
and by the Norwegian Research Council (NFR) under Grant No. 255253/F50 - 
``CERN Heavy Ion Theory." M.T., O.P., and O.V. acknowledge financial 
support of the Norwegian Centre for International Cooperation in 
Education (SIU) under Grant ``CPEA-LT-2016/10094 - From Strong 
Interacting Matter to Dark Matter."
This work was also performed within the European network
COST Action CA15213 ``Theory of hot matter and relativistic heavy-ion
collisions" (THOR). All computer calculations were made at Abel (UiO, 
Oslo) and Govorun (JINR, Dubna) computer cluster facilities.
\end{acknowledgments}

\end{document}